\documentstyle[12pt,epsfig,psfig]{article}
\def\reference{\parskip 0pt\par\noindent\hangindent 0.5 truecm}

\begin{document}
%
% Title
% Capitalise the title normally - do not use ALL CAPS.
%
\title{Ultra High Energy Cosmic Rays}
%

% Authors
% Here comes the author(s) of the paper. Please add the appropriate author
% names for your paper and indicate within the $^...$ the number(s)
% which corresponds to the institute(s) of each author. In this example
% the second author has two institutional affiliations.
% Add or remove authors as required.
% **** IMPORTANT: Leave the closing curly bracket line as is. ******

\author{ R.J. Protheroe$^{1}$ \and R.W. Clay$^{2}$
} % IMPORTANT: leave this curly bracket as the first character of this line.

% Date - leave this blank.
\date{}
\maketitle

% Institutions
% Here fill in your institute name(s) and address(es)
% The number in $^...$ indicates the author number.  For example
{\center Department of Physics, The University of Adelaide\\
Adelaide, Australia 5005\\$^1$rprother@physics.adelaide.edu.au,
$^2$rclay@physics.adelaide.edu.au\\[3mm] }

%{\bf Version 28/9/2003 }

% Abstract
\begin{abstract}
Cosmic rays with energies above $10^{18}$~eV are currently of
considerable interest in astrophysics and are to be further
studied in a number of projects which are either currently under
construction or the subject of well-developed proposals.  This
paper aims to discuss some of the physics of such particles in
terms of current knowledge and information from particle
astrophysics at other energies.
\end{abstract}

{\bf Keywords: cosmic rays, acceleration of particles, magnetic
fields, dark matter, radiation mechanisms: nonthermal, diffuse
radiation}
% Place keywords here. Please write all keywords in lower case. PASA uses the
 %standard list of subject 
% headings adopted by The Astrophysical Journal and available from URL:
%   http://www.journals.uchicago.edu/ApJ/keywords_text.html

% A formatting command to add space between the author list and the body
% of the paper when printed. This spacing may be changed as desired.
\bigskip

%
% Body of paper
%

\section{Introduction}

Cosmic rays (CR) are the non-photon particles with which we learn
about astrophysics.  Their composition ranges over the known
nuclei (and antiprotons) to electrons (and positrons) and
neutrinos of all flavours, and (perhaps) to exotic particles as
yet unobserved in accelerator physics. Cosmic ray studies are
complementary to photon astrophysics since many astrophysical
photons are produced in processes, such as synchrotron emission,
which involve charged cosmic ray particles.  In this review we
concentrate on the highest energy cosmic rays, which are unaffected by
heliospheric modulation, but are strongly affected by propagation
effects through our galaxy. In the case of an extragalacic
origin for the highest energy cosmic rays, their propagation through
intergalactic photon and magnetic fields will have a profound influence on
what we observe.

Radio astronomical studies probe cosmic rays at their sources
and we are led to associate cosmic ray acceleration with
energetic radio objects, pulsars, supernovae, AGNs etc.  The next
generation of large radio projects (LOFAR, SKA etc.) will
certainly add considerably to our understanding of cosmic ray
sources, and may even offer opportunities to develop new
detection techniques through the radio-frequency fields of the
air showers of cosmic rays arriving at Earth (e.g.\ Falcke and Gorham
2003).

The radio emission is dominated by cosmic
ray electrons which rapidly lose energy in the emission processes.  The
more
massive nuclei do not have a clear radio signature but propagate with
little energy loss and are readily observed
at Earth.  They are believed to fill (at least) the volume of our galaxy.
However, being charged particles, their paths are dictated by
astrophysical
magnetic fields and only at the highest energies do we expect magnetic
deflections to be sufficiently small for directional astrophysics to be
possible.

At modest energies (up to about $10^{14}$eV), cosmic rays are
sufficiently plentiful that balloon experiments can have enough
collecting area to directly study properties of the beam.
Indeed, even satellite experiments extend up to 2~TeV/amu (for
the CRN experiment on Spacelab 2, M\"uller et al 1991) which for
iron is about $10^{14}$eV per nucleus.  At higher energies, we
depend on techniques which investigate the cascades of particles
(extensive air showers) resulting from the impact of cosmic ray
particles on our atmospheric gas.  Since a single cosmic ray
particle can produce millions (or billions) of secondary
particles in this way, and those particles scatter in the
atmosphere to hundreds of metres from the original cosmic ray
trajectory, such studies offer uniquely efficient opportunities
for studying fluxes down to levels of primary CR particles per
square kilometre per century.

The cosmic ray cascades may be directly detected with groups of
spaced large-area radiation detectors on the ground (ground
arrays) or indirectly through collecting (with large telescopes)
the photons which they produce, fluorescence light from
atmospheric nitrogen or Cerenkov emission from the bulk
atmospheric gas.  In the future, such light collectors may well
be satellite-based.

The Ultra High Energy cosmic rays (UHE CR) which are the focus
of this review are currently of particular interest.  Spectral
data at energies above 1 EeV
($10^{18}$eV) and directional results, notably from the AGASA
project, are very suggestive of fascinating, unexpected physics
(Hayashida et al.\ 1999).  Furthermore, this field of research is
experimentally challenging and there is controversy in
discrepancies between experimental data from experiments
currently operational or recently discontinued.  A new era in the
field is beginning with the commissioning of the Pierre Auger
Observatory (Auger Collaboration 2001) which comprises a pair of
3000~km$^2$ arrays (one under construction in Mendoza Province,
Argentina, and one planned for Utah, U.S.A) employing both
particle and optical detectors, followed by large-scale Japanese
projects, and possibly the space-based Extreme Universe Space
Observatory project (EUSO is a Europe/Japan/US
collaboration currently under Phase A study as an ESA mission
with the goal of a three-year mission on the ISS starting in
2009, see http://www.euso-mission.org/).

At the present time, our ideas concerning cosmic rays at these
highest energies are predominantly based on the idea that such
particles are likely to be of extragalactic origin.  Their
sources are presumed to be found in some extreme astrophysical
environment such as the most energetic radio or gamma-ray
sources.  We have been forced to these ideas by our failure to
identify any models for galactic objects capable of accelerating
particles to within one thousandth of the required energies, and
also by the failure of our galactic source models to reproduce
the directional isotropy of the observed beam.  Nonetheless,
extragalactic scenarios have their own problems.  The
intergalactic magnetic field has largely unknown properties.  It
could be strong enough, with a structure which makes it
impenetrable to particles from nearby clusters of galaxies in
realistic periods of time.  Also, intracluster magnetic fields
may limit the ability of particles even to leave galactic
clusters.

This paper begins by presenting a brief overview of the observed
properties (energy spectrum, arrival directions, and composition)
of the cosmic ray beam.  There follows a physical discussion of
the cosmic ray acceleration process and resulting limits to the
maximum achievable energies.  Energy losses in the following
propagation through astrophysical fields are fundamental in
determining the observed beam properties and these are then
discussed.  It is possible that the origin of the highest energy
cosmic rays lies in exotic particles and that field of research
is introduced.  Whatever the sources of these particles may be,
their observed directional properties will be closely linked with
the properties of the intervening magnetic fields.  A discussion
of this key factor completes our review.  For an earlier review
see Nagano and Watson (2000).

%%%%%%%%%%%%%%%%%%%%%%%%%%%%%%%%%%%%%%%%%%%%%%%%%%%%%%%%%%%%%%%%%%%%%%%%%%%%%

\subsection{The Detection of Ultra High Energy Cosmic Rays}

Cosmic ray observatories record the arrival directions and
distribution of energies of incident particles.  They also
attempt to measure a remaining parameter, which is the mass
composition of the primary cosmic ray particle.  It is in the
latter respect that cosmic ray studies differ from others in
astrophysics.  The mass composition is a key parameter in our
astrophysical understanding since the charge on the particle
(closely related to mass, for a nucleus), with the momentum (or
energy) determines the propagation path of the particle.

At the highest energies, the cosmic ray flux is extremely low and
the particle recording must be through processes which enable
detection to be achieved at large distances from the path of the
primary cosmic ray.  In practice, this involves making use of the
particle cascade which results from the interaction of the
primary cosmic ray particle with our atmosphere.

That cascade is initiated by a primary cosmic ray particle when
its first atmospheric interaction occurs.  A cascade of secondary
particles is then fed by degrading energy from the primary
particle as it repeatedly interacts in its atmospheric passage.
Those secondary particles cause energy to be deposited into the
atmospheric gas, with some remaining energy reaching the ground.
The cosmic ray detection process then consists in either a
measurement of the passage of the energy as it is carried by
particles through the atmosphere (through any emitted Cerenkov
light or by any induced nitrogen fluorescence light) or in 
direct detections in radiation detectors of remaining particles
which reach the ground.

The arrival direction of the primary particle is deduced from the
direction of the path of the atmospheric cascade and
characteristically has a resolution of a fraction of a degree.
The primary particle energy is deduced by attempting to make
sufficient measurements on the cascade such that a calorimetric
accounting (to a few tens of percent) can be made of the various
energy sinks.  This is most directly done through nitrogen
fluorescence studies, but the signal is weak and the technique
works best at the highest energies.  Mass composition
measurements are the most difficult, and cannot be made on an
individual cascade basis.  They depend on statistical studies of
the cascade developments as a function of energy.  Crudely
speaking, massive nuclei have a large cross section and interact
early.  Their cascades also degrade in energy relatively rapidly.
As a result, early developing cascades are signatures of 'heavy'
primaries and late developing cascades are indicative of 'light'
(probably proton) primaries.  Massive amounts of cascade
modelling puts flesh on these arguments.

Nitrogen fluorescence studies were pioneered by the Fly's Eye
experiment (Baltrusaitis et al.\ 1985) and its
successor, the High Resolution Fly's Eye (HiRes).  Many ground
detector arrays have been used.  The largest one in fully
developed use is the AGASA (Hayashida et al.\
1999) experiment in Japan.  Both of the HiRes
and AGASA experiments are sufficiently large to probe energies in
excess of $10^{20}$eV.  The next of these huge experiments will
use both techniques.  The Pierre Auger Observatory (Auger
Collaboration 2001) is designed to operate above
$10^{19}$eV.  It will have a 3000 square kilometre collecting
area instrumented with 1600 large-area particle detectors and
will also have twenty-four 4~m diameter Schmidt optical telescopes
for fluorescence detection.

\subsection{The Cosmic Ray Energy Spectrum}

\begin{figure}
\epsfig{file=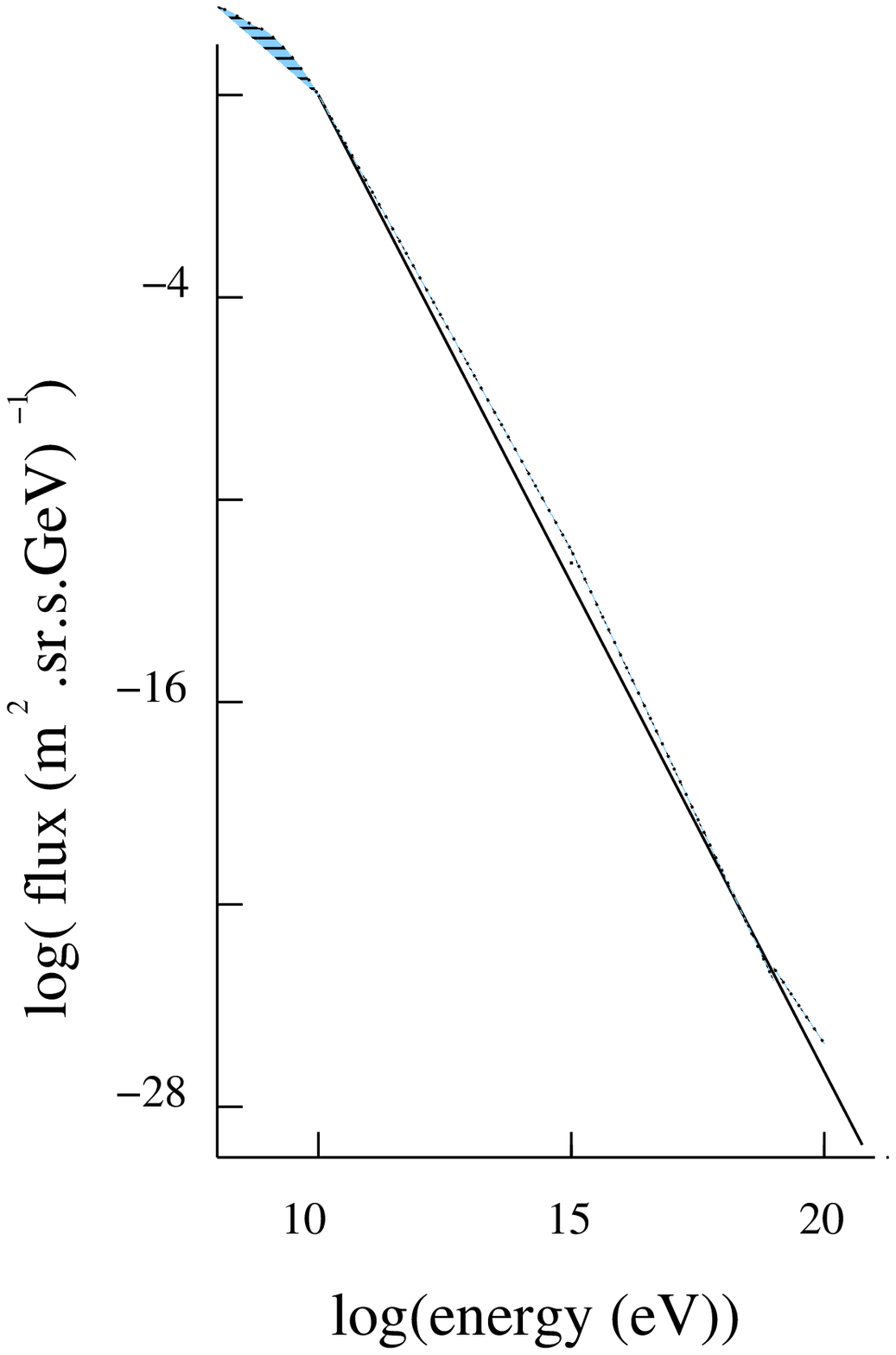,width=120mm}
\caption{ The cosmic ray energy spectrum (dotted) as measured from the Earth
(after Bhattacharjee and Sigl 2000).}
\label{fig:spectrum}
\end{figure}

The cosmic ray energy spectrum is shown in
Fig.~\ref{fig:spectrum}.  It is remarkable both in its range of
energies and in its range of fluxes.  It covers over ten decades
of energy and thirty decades of flux in a form close to a power
law with an index of about -2.7.  Deviations from that power law
are relatively small but are generally regarded as physically
significant.  There is a steepening at about $10^{16}$eV, known
as the knee, and a flattening at about $10^{18}$eV known as the
ankle (e.g.\ Abu-Zayyad et al.\ 2001).  The knee is
often argued to be associated with an energy limit of
acceleration from supernova remnant sources, although it may well
be related to a loss of ability for our galactic magnetic fields
to retain (and build up internally) the cosmic ray flux (e.g.\ Clay
2002).  The ankle is usually associated with the
onset of a dominant, flatter, extra-galactic cosmic ray
spectrum. It is important to note that, in this model, our galaxy
produces particles with energies up to those of the ankle of the
spectrum.  That is already above 1~EeV$\equiv$$10^{18}$eV, and is well above
energies which are easily accessible for any present galactic
acceleration models.

A key region of the spectrum is its very highest energies.  The
flux here is so low that the low event statistics in our
observations to date leave us uncertain of the spectral structure
above the key energy of $6\times 10^{19}$eV.  Here there is a
predicted spectral downturn, the Greisen-Zatsepin-Kuzmin (GZK)
cut-off (Greisen 1966, Zatsepin and Kuzmin 1966) for particles
which have travelled more than a few tens of Mpc, due to
interactions with the 2.7K cosmic microwave background radiation
(CMBR).  However, several experiments have reported CR events
with energies above $10^{20}$~eV (Takeda et al.\ 2003) with the
highest energy event having 300~EeV (Bird et al.\ 1995).  Very
recent data from the two largest aperture high energy cosmic ray
detectors are contradictory: AGASA (Takeda et al.\ 2003) observes
no GZK cut-off while HiRes (Abu-Zayyad et al. 2002) observes a
cut-off consistent with the expected GZK cut-off.  A systematic
over-estimation of energy of about 25\% by AGASA or
under-estimation of energy of about 25\% by HiRes could account
the discrepancy (Abu-Zayyad et al. 2002), but the continuation of
the UHECR spectrum to energies well above $10^{20}$ eV is now far
from certain.  Future measurements with Auger (Auger
Collaboration 2001) should resolve this question.  If the
spectrum does extend well beyond $10^{20}$ eV, determining the
origin of these particles could have important implications for
astrophysics, cosmology and particle physics.

%%%%%%%%%%%%%%%%%%%%%%%%%%%%%%%%%%%%%%%%%%%%%%%%%%%%%%%%%%%%%%%%%%%%%%%%%%%%%

\subsection{Arrival Directions}

A key observation in cosmic ray astrophysics is the directional
distribution of the particles.  That distribution will depend on
any galactic magnetic fields and hence will be energy (rigidity)
dependent.  However, with very limited exceptions, which are not
individually statistically significant, there is no observed
deviation from isotropy above the knee of the energy spectrum,
and any anisotropies at lower energies are themselves very small
(Smith and Clay 1997, Clay, McDonough and Smith 1997).  Recently,
the AGASA experiment (Hayashida et al.\ 1999) found a
non-uniform distribution of arrival directions, suggestive of a
source direction, in the energy range $10^{18.0}$eV to
$10^{18.4}$eV.  That observation is potentially very important,
particularly as there is supporting evidence in data from the
SUGAR array (Bellido et al.\ 2001).  However, neither of those
observations on their own is clearly statistically significant.
Still, those data are regarded by many as the possible beginning
of a new era in cosmic ray astrophysics in which we can begin
directional cosmic ray astronomy.  The possibility of having a
source to observe may indeed open up new frontiers.

%%%%%%%%%%%%%%%%%%%%%%%%%%%%%%%%%%%%%%%%%%%%%%%%%%%%%%%%%%%%%%%%%%%%%%%%%%%%%

\subsection{Composition of the UHECR}

The highest energy cosmic rays show no major differences in their
air shower characteristics to cosmic rays at lower energies.  One
would therefore expect the highest energy cosmic rays to be
protons particularly if, as is most likely, they are
extragalactic in origin.  However, it is still possible that they
are not single nucleons.  Obvious candidates are heavier nuclei
(e.g. Fe), $\gamma$-rays and neutrinos.  Surprisingly, it is even
more difficult to propagate nuclei than protons, because of the
additional photonuclear disintegration (Tkaczyk et al. 1975,
Puget et al. 1976, Karakula and Tkaczyk 1993, Elbert and Sommers
1995, Anchordoqui et al.\ 1997, Stecker and Salamon
1999).
The possibility that the 300 EeV Fly's Eye event is a
$\gamma$-ray has been discussed (Halzen et al. 1995)
and, although not completely ruled out, the air shower
development profile seems inconsistent with a $\gamma$-ray
primary.  Weakly interacting particles such as neutrinos will
have no difficulty in propagating over extragalactic distances,
of course.  This possibility has been considered, and generally
discounted (Halzen et al. 1995, Elbert and Sommers
1995), mainly because of the relative
unlikelihood of a neutrino interacting in the atmosphere, and the
resulting great increase in the luminosity required of cosmic
sources.

%%%%%%%%%%%%%%%%%%%%%%%%%%%%%%%%%%%%%%%%%%%%%%%%%%%%%%%%%%%%%%%%%%%%%%%%%%%%%

\subsection{Cosmic Ray Sources}

\begin{figure}
\epsfig{file=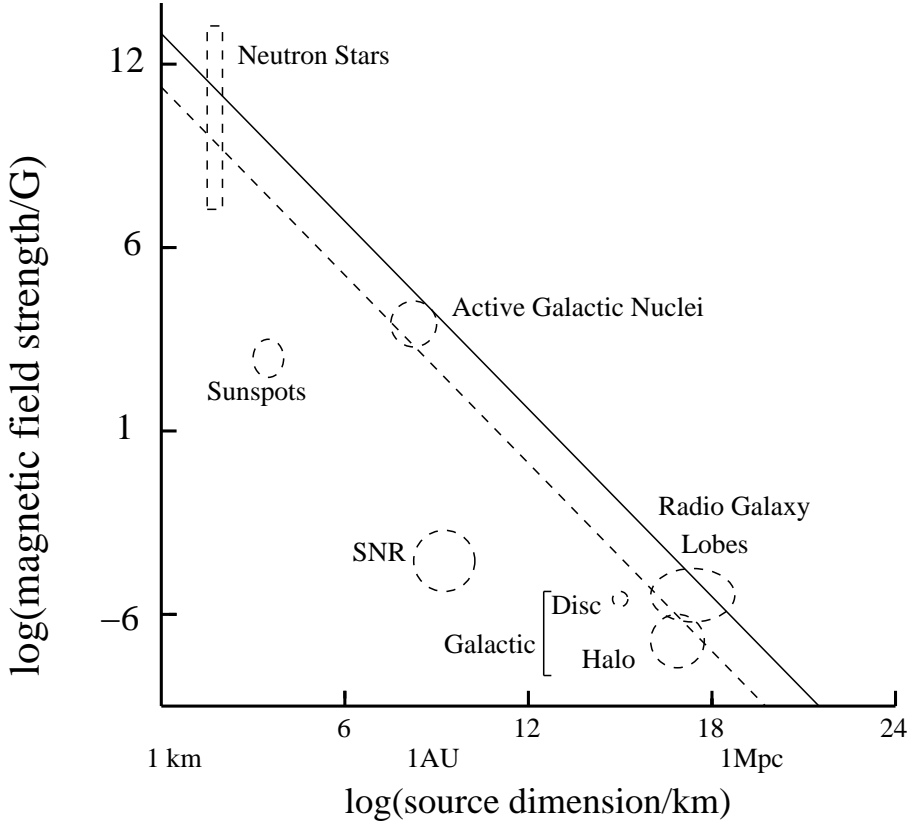,width=120mm}
\caption{ Proposed sites for cosmic ray acceleration related to
their likely dimensions and magnetic field strength for
scattering centres moving with $v_{\rm scat}\sim c$ (after Hillas
1984).  The lines represent plausible limits for $10^{20}$eV
cosmic ray containment in the sources (solid line -- protons,
dashed line -- iron nuclei) for scattering centres moving at at
speeds $\sim c$.  For scattering centres moving at speeds $v_{\rm
scat} < c$ the size would need to be a factor $c/v_{\rm scat}$
larger.}
\label{fig:hillas_plot}
\end{figure}

Figure~\ref{fig:hillas_plot} is a well known diagram first
produced by Hillas (1984).  It reminds us that acceleration,
associated with magnetic structures, requires the field and its
dimensions to be sufficient to contain the accelerating particle
through the acceleration process.  The lines simply give the
magnetic field vs.\ gyroradii $r_g$ for protons (solid) and iron
nuclei (dashed), and this gives the minimum size for scattering
centres moving at at speeds $\sim c$.  More realistically, for
scattering centres moving at speeds $v_{\rm scat} < c$ the size
would need to be a factor $c/v_{\rm scat}$ larger.  This puts a
limit on the product of the source field and its physical
dimensions.  Strong fields with large-scale structure are
attractive for acceleration to the highest energies.  The
acceleration is thought to be likely to be associated with
astrophysical shocks.

One of the very few plausible
acceleration sites may be associated with the radio lobes of
powerful radio galaxies, either in the hot spots (Rachen and
Biermann 1993) or possibly the cocoon
or jet (Norman et al. 1995).  One-shot processes
such as magnetic reconnection (e.g. in jets or accretion disks)
comprise another possible class of sources (Haswell et al. 1992,
Sorrell 1987).

Acceleration at the termination shock of the galactic wind from
our Galaxy has also been suggested by Jokipii and Morfill
(1985), but due to the lack of any statistically
significant anisotropy associated with the Galaxy it is unlikely
to be the explanation.  However, a re-evaluation of the
world data set of cosmic rays has shown that there is a
correlation of the arrival directions of cosmic rays above 40 EeV
with the supergalactic plane (Stanev et al. 1995),
lending support to an extragalactic origin above this energy, and
in particular to models where ``local'' sources ($<100$ Mpc)
would appear to cluster near the supergalactic plane.  Such a
correlation would also be consistent with a Gamma Ray Burst (GRB)
origin as two type Ic hypernovae (supernovae with broad
absorption features indicating high velocity ejected material and
a rather large explosion energy) have now been identified with
GRB (SN 2003dh/GRB 030329 Kawabata et al.\ 2003;
SN 1998bw/GRB980425 Galama et al.\ 1998).  The
expanding fireball may have ultrarelativistic components (e.g.\
$\Gamma \sim 300$) and this may lead to production of UHECR
through relativistic shock acceleration (Vietri
1995) or some other process (see e.g.\ Dermer 2002
for a discussion and references to earlier work).

Because of the resulting flat spectrum of particles (including
gamma-rays and protons) extending up to GUT (grand unified
theory) scale energies, topological defect models
have been invoked to
try to explain the UHE CR.  Propagation of the spectra of all
particle species over cosmological distances is necessary to
predict the cosmic ray and gamma-ray spectra expected at Earth.
Propagation over cosmological distances to Earth (as would be the
case in some topological defect origin models) results in
potentially observable gamma-ray fluxes at GeV energies in
addition to cosmic rays.  Massive relic particles on the other
hand, would cluster in galaxy halos, including that of our
Galaxy, and may give rise to anisotropic cosmic ray signals at
ultra high energies.

The suggestion that neutron stars might accelerate cosmic rays
followed soon after the discovery of pulsars (see, e.g.\ Gold
1975 and references therein).  Voltages up to $\sim
10^{12}$--$10^{15}$V (depending on pulsar period and magnetic
field) are expected in a pulsar's magnetosphere.  These could
accelerate nuclei with a resulting flat spectrum extending up to
$\sim$10$^{16}$~eV, and could possibly explain the knee in the
cosmic ray spectrum (e.g.\ Bednarek \& Protheroe 2002).  The
pulsar wind shock has been proposed as an acceleration site
(e.g., Berezhko 1994, Bell and Lucek 1996) and might in principle
accelerate particles to $10^{15}$--$10^{19}$eV.  Blasi et al.\
(2000) suggest that the UHE CR events are due to iron nuclei
accelerated from young strongly magnetized neutron stars through
relativistic MHD winds of neutron stars whose initial spin
periods are shorter than $\sim 10$ ms.  More recently, the
acceleration to ultra-high energies has been discussed in the
context of fast re-connection in millisecond pulsars, but it
appears that to reach $\sim 10^{20}$~eV magnetic fields $\sim
10^{12}$--$10^{15}$~G, and special geometries are required (de
Gouveia Dal Pino and Lazarian, 2001).  In our opinion, while
neutron stars may contribute to the observed cosmic ray spectrum
it seems unlikely that they are responsible for the UHE CR, with
the possible exception of transient shocks in pulsar winds of
neutron stars formed during a hypernova explosion (i.e. a GRB as
already discussed).

We shall delay a more detailed discussion of radio galaxies,
active galactic nuclei and topological defects as sources of the
UHE CR, and discuss next diffusive shock acceleration and cosmic
ray propagation.

%%%%%%%%%%%%%%%%%%%%%%%%%%%%%%%%%%%%%%%%%%%%%%%%%%%%%%%%%%%%%%%%%%%%%%%%%%%%%

\section{Diffusive Shock Acceleration}

For stochastic particle acceleration by electric fields induced by the
motion of magnetic fields $B$, the rate of energy gain by relativistic
particles of charge $Ze$ can be written (in SI units)
\begin{equation}
\left. {dE \over dt} \right|_{\rm acc} = \xi(E) Ze c^2 B
\label{eq:max_acc}
\end{equation}
where $\xi(E) < 1$ and depends on the acceleration mechanism.
Below is a simple heuristic treatment of Fermi acceleration based
on those given by Gaisser (1990) and Protheroe
(2000).  We shall start with 2nd
order Fermi acceleration (Fermi's original theory) and describe
how this can be modified in the context of astrophysical shocks
into the more efficient 1st order Fermi mechanism known as
diffusive shock acceleration.  More detailed and rigorous
treatments are given in several review articles (Drury 1983a,
Blandford and Eichler 1987, Berezhko and Krymsky
1988).  See
the review by Jones and Ellison (1991) on
the plasma physics of shock acceleration which also includes a
brief historical review and refers to early work.

\subsection{Fermi's Original Theory}

Gas clouds in the interstellar medium have random velocities of
$\sim 15$ km/s superimposed on their regular motion around the
galaxy.  Cosmic rays gain energy on average when scattering off
these magnetized clouds.  A cosmic ray enters a cloud and
scatters off irregularities in the magnetic field which is tied
to the partly ionized cloud.

\begin{figure}[htb]
\epsfig{file=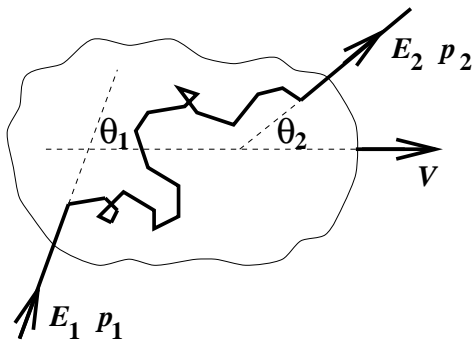,height=4.5cm}
\caption{Interaction of cosmic ray of energy $E_1$ with ``cloud''
moving with speed $V$
\label{fig:fermi_acc_orig}}
\end{figure}

In the frame of the cloud: (a) there is no change in energy
because the scattering is collisionless, and so there is elastic
scattering between the cosmic ray and the cloud as a whole which
is much more massive than the cosmic ray; (b) the cosmic ray's
direction is randomized by the scattering and it emerges from the
cloud in a random direction.

Consider an ultra-relativistic cosmic ray entering a cloud with energy $E_{1}$ and
momentum $p_{1}\approx E_1/c$ travelling in a direction making angle
$\theta_{1}$ with the cloud's direction.  After scattering inside
the cloud, it emerges with energy $E_{2}$ and momentum $p_{2}\approx E_2/c$ at
angle $\theta_{2}$ to the cloud's direction
(Fig.~\ref{fig:fermi_acc_orig}).  The energy change is obtained
by applying the Lorentz transformations between the laboratory
frame (unprimed) and the cloud frame (primed).  Transforming to
the cloud frame:
\begin{equation}
E_{1}^{\prime} = \gamma E_{1} (1 - \beta \cos \theta_{1})
\end{equation}
where $\beta = V/c$ and $\gamma = 1/\sqrt{1-\beta^{2}}$.

\noindent Transforming to the laboratory frame:
\begin{equation}
E_{2} = \gamma E_{2}^{\prime} (1 + \beta \cos \theta_{2}^{\prime}).
\end{equation}

The scattering is collisionless, being with the magnetic field.
Since the magnetic field is tied to the cloud, and the cloud is
very massive, in the cloud's rest frame there is no change in
energy, $E_{2}^{\prime} = E_{1}^{\prime}$, and hence we obtain
the fractional change in LAB-frame energy $(E_{2}-E_{1})/E_{1}$,
\begin{equation}
{\Delta E \over E} = 
{1 - \beta \cos \theta_{1} + \beta \cos \theta_{2}^{\prime}
- \beta^{2} \cos \theta_{1} \cos \theta_{2}^{\prime} \over
1 - \beta^{2}} -1.
\label{eq:egain_general}
\end{equation}

We need to obtain average values of $ \cos \theta_{1}$ and $ \cos
\theta_{2}^{\prime}$.  Inside the cloud, the cosmic ray scatters
off magnetic irregularities many times so that its direction is
randomized,
\begin{equation}
\langle \cos \theta_{2}^{\prime} \rangle =0.
\end{equation}
The average value of cos$\theta_{1}$ depends on the rate at which
cosmic rays collide with clouds at different angles.  The rate of
collision is proportional to the relative velocity between the
cloud and the particle so that the probability per unit solid
angle of having a collision at angle $\theta_{1}$ is proportional
to $(v - V \cos \theta_{1})$.  Hence, for ultrarelativistic
particles ($v=c$)
\begin{equation}
{dP \over d \Omega_{1}} \propto (1 - \beta \cos \theta_{1}),
\end{equation}
and we obtain
\begin{equation}
\langle \cos \theta_{1} \rangle = 
\int \cos \theta_{1} {dP \over d \Omega_{1}} d \Omega_{1} /
\int {dP \over d \Omega_{1}} d \Omega_{1} = - {\beta \over 3},
\end{equation} 
giving
\begin{equation}
{\langle \Delta E \rangle \over E} = 
{1 + \beta^{2}/3 \over
1 - \beta^{2}} -1 \simeq {4 \over 3} \beta^{2}
\end{equation}
if $\beta \ll 1$.

We see that $\langle \Delta E \rangle / E \propto \beta^{2}$ is
positive (energy gain), but is 2nd order in $\beta$ and if $\beta
\ll 1$ the average energy gain per cloud collision is very small.
This is because there are almost as many overtaking collisions
(energy loss) as there are head-on collisions (energy gain).  The
acceleration rate is
\begin{equation}
r_{\rm acc}(E) \equiv {1 \over E} \left. {dE \over dt}
\right|_{\rm acc} = {\langle \Delta E \rangle \over E} r_{\rm
coll} \simeq {4 \over 3} \left({V \over c}\right)^2 r_{\rm coll}
\end{equation}
where $r_{\rm coll}$ is the rate of collision of the cosmic ray
with the cloud.  2nd order Fermi acceleration is one example of
stochastic acceleration.  Another example involves the average
energy gain that arises in resonant interactions of cosmic rays
with Alfv\'en waves.  In this case, the acceleration rate would
be
\begin{equation}
r_{\rm acc}(E)  \propto  \left({v_A \over v}\right)^2  r_{\rm coll}
\label{eq:stochastic}
\end{equation}
where $v_A$ is the Alfv\'en velocity and $r_{\rm coll}$ is now
the rate of collision of the cosmic ray (speed $v$) with the
Alfv\'en waves (see e.g.\ Jones 1994).

%%%%%%%%%%%%%%%%%%%%%%%%%%%%%%%%%%%%%%%%%%%%%%%%%%%%%%%%%%%%%%%%%%%%%%%%%%%%

\subsection{1st Order Fermi Acceleration at Astrophysical Shocks}

Fermi's original theory was modified in the 1970's (Axford, Lear
and Skadron 1977, Krymsky 1977, Bell 1978, Blandford and Ostriker
1978) to describe more efficient acceleration (1st order in
$\beta$) taking place at supernova shocks but is generally
applicable to strong shocks in other astrophysical contexts.  Our
discussion of shock acceleration will be of necessity brief, and
omit a number of subtleties.

\begin{figure}[htb]
\epsfig{file=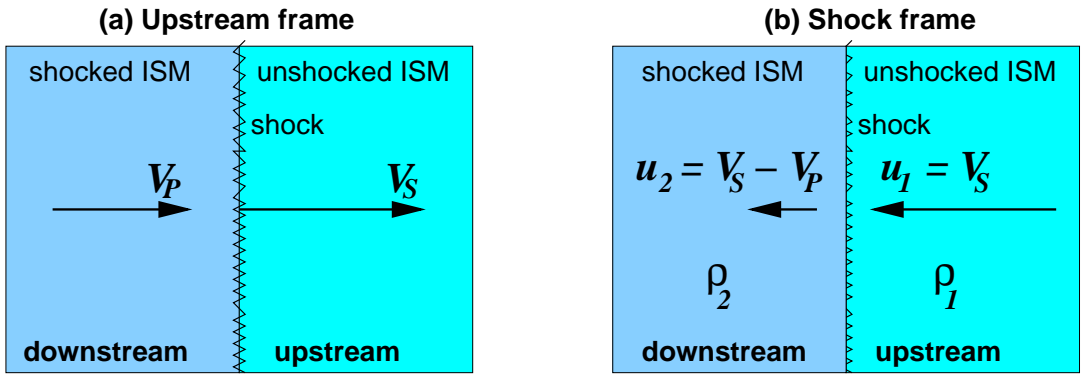,height=5cm}
\caption{A shock moving with speed $V_S$ ahead of shocked gas
moving at the pistion speed $V_P$ as seen (a) in the upstream
frame, (b) in the shock frame.}
\label{fig:shock_vel}
\end{figure}

Here, for simplicity, we adopt the test particle approach
(neglecting effects of cosmic ray pressure on the shock profile),
adopt a plane geometry and consider only non-relativistic shocks.
Nevertheless, the basic concepts will be described in sufficient
detail that we can consider acceleration and interactions of the
highest energy cosmic rays, and to what energies they can be
accelerated.  We consider the classic example of a SN shock.
During a supernova explosion several solar masses of material are
ejected at a speed of $\sim 10^{4}$ km/s which is much faster
than the speed of sound in the interstellar medium (ISM) which is
$\sim$ 10 km/s.  A strong shock wave propagates radially out
(speed $V_{S}$) as the ISM and its associated magnetic field
piles up in front of the supernova ejecta which moves at speed
$V_{P}$ (see Fig.~\ref{fig:shock_vel}a).  As seen in the frame of
the shock (see Fig.~\ref{fig:shock_vel}b) gas flows from upstream
into the shock with speed $u_1=V_{S}$ and density $\rho_1$, and
flows out of the shock downstream with speed $u_2=(V_S-V_P)$ and
density $\rho_2$.  The velocity of the shock, $V_{S}$, depends on
the velocity of the piston (ejecta), $V_{P}$, and on the ratio of
specific heats, $\gamma$.  The compression ratio for a strong
non-relativistic shock is given by
\begin{equation}
R = {\rho_2 \over \rho_1} = {u_1 \over u_2} = {\gamma + 1 \over \gamma - 1 }
\end{equation}
from which $V_{S}=Ru_2$ and  $V_P=(u_1-u_2)=(R-1)u_2$ giving
\begin{equation}
V_{S}/V_{P} \simeq R/(R-1).
\end{equation}
For SN shocks the SN will have ionized the surrounding gas which
will therefore be monatomic ($\gamma = 5/3$), and so a strong
shock will have $R=4$.

\begin{figure}[htb]
\epsfig{file=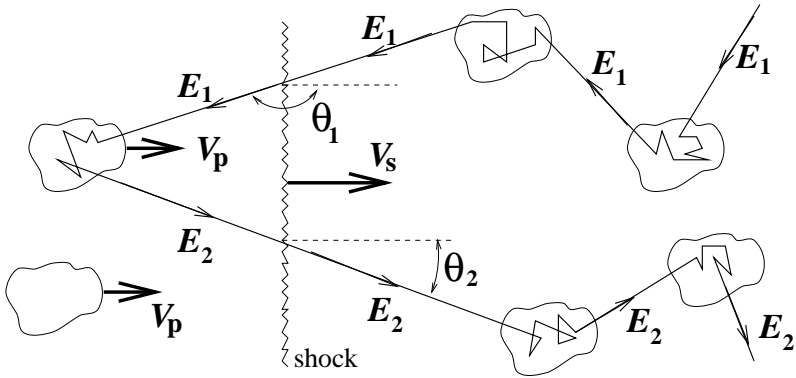,height=5cm}
\caption{Interaction of cosmic ray of energy $E_1$ with a shock
moving with speed $V_s$.
\label{fig:fermi_acc_shock}}
\end{figure}

In order to work out the energy gain per shock crossing, we can
visualize magnetic irregularities tied to the plasma on either
side of the shock as clouds of magnetized plasma of Fermi's
original theory (Fig.~\ref{fig:fermi_acc_shock}).  Here, we
assume that the shock is non-relativistic, such that we can make
the approximation that the ultra-relativistic accelerated
particles are isotropic in both upstream and downstream frames.
By considering the rate at which cosmic rays cross the shock from
downstream to upstream, and upstream to downstream, one finds
$\langle \cos \theta_{1} \rangle \approx -2/3$ and $\langle \cos
\theta_{2}^{\prime} \rangle \approx 2/3$, and
Eq.~\ref{eq:egain_general} then gives
\begin{equation}
{\langle \Delta E \rangle \over E}  \simeq {4 \over 3} \beta 
\simeq {4 \over 3} {V_{P} \over c} 
\simeq {4 \over 3} {(R-1) \over R} {V_{S} \over c}.
\end{equation}
Note this is 1st order in $\beta=V_{P}/c$, and so the fractional
energy change per collision can be much higher than in Fermi's
original theory.  This is because of the converging flow --
whichever side of the shock you are on, if you are moving with
the plasma, the plasma on the other side of the shock is
approaching you at speed $V_p$.

\begin{figure}[htb]
\epsfig{file=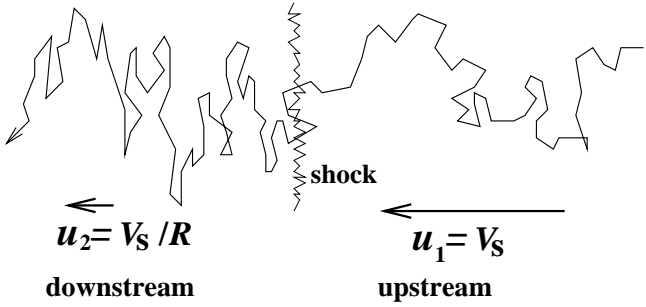,height=4cm}
\caption{Diffusion of cosmic rays from upstream to downstream
seen in the shock frame.
\label{fig:up_downstream}}
\end{figure}

To obtain the energy spectrum we need to find the probability of
a cosmic ray encountering the shock once, twice, three times,
etc.  If we look at the diffusion of a cosmic ray as seen in the
rest frame of the shock (Fig.~\ref{fig:up_downstream}), there is
clearly a net flow of the energetic particle population in the
downstream direction.

The flux of cosmic rays lost downstream is
\begin{equation}
f_{{\rm loss}} = n_{\rm CR} V_{S}/R 
\label{eq:r_loss}
\end{equation}
since cosmic rays with number density $n_{\rm CR}$ at the shock
are advected downstream with speed $u_2=V_{S}/R$ (from right to left
in Fig.~\ref{fig:up_downstream}) and we have neglected
relativistic transformations of the rates because $V_S \ll c$.
 
Upstream of the shock, cosmic rays travelling at speed $v$ at
angle $\theta$ to the shock normal (as seen in the laboratory
frame) approach the shock with speed $(V_{S} + v \cos \theta)$ as
seen in the shock frame.  Clearly, to cross the shock, $\cos
\theta > -V_{S}/v$.  Then, assuming cosmic rays upstream are
isotropic, the flux of cosmic rays  crossing from upstream to
downstream is
\begin{eqnarray}
f_{\rm cross} &=&  n_{\rm CR} {1 \over 4 \pi} \int_{-V_S/v}^1 
(V_S +  v \cos \theta) 2 \pi d( \cos \theta) \;  \approx \; n_{\rm CR} v/4 .
\label{eq:r_cross}
\end{eqnarray}

The probability of crossing the shock once and then escaping from
the shock (being lost downstream) is the ratio of these two
fluxes:
\begin{equation}
{\rm Prob.(escape)} = f_{\rm loss}/f_{\rm cross} \approx 4 V_{S}/Rv.
\end{equation}
The probability of returning to the shock
after crossing from upstream to downstream is
\begin{equation}
{\rm Prob.(return)} = 1 - {\rm Prob.(escape)},
\end{equation}
and so the probability of returning to the shock $m$ times and
also of crossing the shock at least $m$ times is
\begin{equation}
{\rm Prob.(cross} \ge m{\rm )} = [1 - {\rm Prob.(escape)}]^{m}.
\end{equation}
The energy after $m$ shock crossings is
\begin{equation}
E = E_{0} \left( 1 + {\Delta E \over E} \right)^{m}
\end{equation}
where $E_{0}$ is the initial energy.

To derive the spectrum, we note that the integral energy spectrum
(number of particles with energy greater than $E$) on
acceleration must be
\begin{equation}
N(>E) \propto  [1 - {\rm Prob.(escape)}]^{m}
\end{equation}
where 
\begin{equation}
m = {\ln (E/E_{0}) \over \ln (1 + \Delta E/E)}.
\end{equation}
Hence,
\begin{equation}
\ln N(>E) = A + {\ln (E/E_{0}) \over \ln (1 + \Delta E/E)} 
\ln [1 - {\rm Prob.(escape)}],
\end{equation}
where $A$ is a constant, and so
\begin{equation}
\ln N(>E) = B - (\Gamma-1) \ln E
\end{equation}
where $B$ is a constant and
\begin{equation}
\Gamma = \left(1 - {\ln [1 - {\rm Prob.(escape)}] \over \ln (1 + \Delta
E/E)}\right) \approx {R+2 \over R-1}
\label{eq:gamma}
\end{equation}
where we have used $\ln(1+x) \approx x$ for $x \ll 1$.

Hence we arrive at the spectrum of cosmic rays on acceleration
\begin{equation}
N(>E) \propto E^{{-(\Gamma-1)}} \hspace{1cm} \rm (integral \; form)
\end{equation}
\begin{equation}
{dN \over dE} \propto E^{{-\Gamma}} \hspace{1cm} \rm (differential \; form).
\end{equation}
For compression ratio $R=4$ (strong shock) we have the well-known
$E^{-2}$ differential spectrum.  The observed cosmic ray spectrum
is steepened by energy-dependent escape of cosmic rays from the
Galaxy.

\subsection{Acceleration Rate}

Here we again neglect effects of cosmic ray pressure and
consider only a non-relativistic shock.  The acceleration rate is defined by
\begin{equation}
r_{\rm acc}(E) \equiv {1 \over E} \left. {dE \over dt} \right|_{\rm
acc} = {(\langle \Delta E \rangle /E) \over t_{\rm cycle}(E)}
\approx {4 \over 3} {(R-1) \over R} {V_{S} \over c \; t_{\rm
cycle}(E)}
\end{equation}
where $t_{\rm cycle}$ is the time for one complete cycle, i.e.
from crossing the shock from upstream to downstream, diffusing
back toward the shock and crossing from downstream to upstream,
and finally returning to the shock. 

The rate of loss of accelerated particles downstream is the probability
of escape per shock crossing divided by the cycle time
\begin{equation}
r_{\rm esc}(E) = {{\rm Prob.(escape)} \over t_{\rm cycle}} \approx
{4 \over R} {V_{S} \over c \; t_{\rm cycle}(E)} = {3 \over R-1}r_{\rm acc}(E)
\end{equation}

We see immediately that the ratio of the escape rate to the
acceleration rate depends on the compression ratio
\begin{equation}
{r_{\rm esc}(E) \over r_{\rm acc}(E)} \approx {3 \over R-1} = (\Gamma-1)
\label{eq:esc_acc}
\end{equation}
and for a strong shock ($R=4$) the two rates are equal, giving
the well-known $E^{-2}$ power-law.

We shall discuss these processes in the shock frame (see
Fig.~\ref{fig:up_downstream}) and consider first particles
crossing the shock from upstream to downstream and diffusing back
to the shock, i.e.\ we shall work out the average time spent
downstream.  Since we are considering non-relativistic shocks,
the time scales are approximately the same whether measured in
the upstream or downstream plasma frame, and so in this section
we drop the use of subscripts indicating the frame of reference.

Diffusion takes place in the presence of advection at speed $u_2$
in the downstream direction.  The diffusion coefficient will be a
function of magnetic rigidity which, for ultra-relativistic
particles considered in this paper, is approximately equal to
$E/Ze$ where $Ze$ is the charge.  However, here we are mainly
concerned with singly charged particles and shall work in terms
of $E$ rather than rigidity.  The diffusion coefficient depends
on the turbulence in the magnetic field.  Often a power-law
dependence $k(E) \propto E^\delta$ is assumed, where $\delta$
depends on the spectrum of turbulence.  For a Kolmogorov spectrum
$\delta=1/3$, and for a completely tangled magnetic field
$\delta=1$ at least over some range of energies.

The typical distance a particle diffuses in time t is
$\sqrt{k_2t}$ where $k_2$ is the diffusion coefficient in the
downstream region.  The distance advected in this time is simply
$u_2t$.  If $\sqrt{k_2t} \gg u_2t$ the particle has a very high
probability of returning to the shock, and if $\sqrt{k_2t} \ll
u_2t$ the particle has a very high probability of never returning
to the shock (i.e. it has effectively escaped downstream).  So,
we set $\sqrt{k_2t} = u_2t$ to define a distance $k_2/u_2$
downstream of the shock which is effectively a boundary between
the region closer to the shock where the particles will usually
return to the shock and the region farther from the shock in
which the particles will usually be advected downstream never to
return.  There are $n_{\rm CR} k_2/u_2$ particles per unit area
of shock between the shock and this boundary.  Dividing this by
$f_{\rm cross}$ (Eq.~\ref{eq:r_cross}) we obtain the average time
spent downstream before returning to the shock
\begin{equation}
t_2(E) \approx {4 \over c} {k_2(E) \over u_2}.
\end{equation}

Consider next the other half of the cycle after the particle has
crossed the shock from downstream to upstream until it returns to
the shock.  In this case we can define a boundary at a distance
$k_1/u_1$ upstream of the shock such that nearly all particles
upstream of this boundary have never encountered the shock, and
nearly all the particles between this boundary and the shock have
diffused there from the shock.  Then dividing the number of
particles per unit area of shock between the shock and this
boundary, $n_{\rm CR} k_1/u_1$, by $f_{\rm cross}$ we obtain the
average time spent upstream before returning to the shock
\begin{equation}
t_1(E) \approx {4 \over c} {k_1(E) \over u_1},
\end{equation}
and hence the cycle time
\begin{equation}
t_{\rm cycle}(E) \approx {4 \over c} \left( {k_1(E) \over u_1} + {k_2(E)
\over u_2} \right).
\end{equation}
The acceleration rate is then given by
\begin{equation}
r_{\rm acc}(E) \approx {(R-1)u_1 \over 3R} \left( {k_1(E) \over u_1} + {k_2(E)
\over u_2} \right)^{-1}.
\label{eq:acc_rate}
\end{equation}

At this point, a comparison with stochastic acceleration is
appropriate.  Noting that the diffusion coefficient can be
written
\begin{equation}
k =  {1 \over 3} \lambda_{\rm coll}v = {1 \over 3}v^2/r_{\rm coll}
\end{equation}
where $\lambda_{\rm coll}$ and $r_{\rm coll}$ are the effective
collision mean free path and collision frequency, respectively,
the acceleration rate can be written
\begin{equation}
r_{\rm acc}(E) \propto \left({u_1 \over v}\right)^2  r_{\rm coll}
\end{equation}
which has the same functional form as for stochastic acceleration
acceleration (Eq.~\ref{eq:stochastic}) as pointed out by Jones
(1994).  

For either shock acceleration or stochastic acceleration to be
able to accelerate cosmic rays to high energies the physical
conditions must be suitable: Alfv\'en waves, magnetic
irregularities or turbulence in the magnetic field must be
present on length scales of the gyroradii of particles
being accelerated and provide a sufficiently high scattering rate
such that the required maximum energy can be achieved during the
life-time of the accelerator, and the large-scale magnetic field
must be able to confine the the highest energy particles within
the accelerator.  There are only a few places (solar flares)
where the turbulence is likely to be strong enough for stochastic
acceleration to work, and the spectrum will differ for different
species.  Generally, shock acceleration is favoured for the
following reasons: (a) collisionless shocks exist everywhere,
provide the necessary physical conditions, and are known to
accelerate particles efficiently; (b) the energy associated with
shocks can be large (a significant fraction of the energy
released in a supernova explosion is carried by the supernova
ejecta); (c) in the test particle case at least, the spectrum of
accelerated particles is a power law which is the same for all
species and depends only on the compression ratio (see
Eq.~\ref{eq:gamma}).

\subsection{Maximum Acceleration Rate}

We next consider the diffusion for the cases of parallel,
oblique, and perpendicular shocks, and estimate the maximum
acceleration rate for these cases.  The diffusion coefficients
required, $k_1$ and $k_2$, are the coefficients for diffusion
parallel to the shock normal.  The diffusion coefficient along
the magnetic field direction is some factor $\eta$ times the
minimum diffusion coefficient, known as the Bohm diffusion
coefficient,
\begin{equation}
k_\parallel = \eta {1 \over 3} r_g c
\end{equation}
where $r_g$ is the gyroradius, and $\eta > 1$.

Parallel shocks are defined such that the shock normal is
parallel to the magnetic field ($\vec{B} || \vec{u_1}$).  In this
case, making the approximation that $k_1 = k_2 = k_\parallel$ and
$B_1 = B_2$ one obtains
\begin{equation}
t_{\rm acc}^\parallel \approx {20 \over 3} {\eta E \over e B_1 u_1^2}.
\end{equation}
For a shock speed of $u_1 = 0.1 c$ and $\eta=10$ one obtains an
acceleration rate (in SI units) of
\begin{equation}
\left. {dE \over dt} \right|_{\rm acc} \approx 1.5 \times 10^{-4}
e c^2 B.
\end{equation}

For the oblique case, the angle between the magnetic field
direction and the shock normal is different in the upstream and
downstream regions, and the direction of the plasma flow also
changes at the shock.  The diffusion coefficient in the direction
at angle $\theta$ to the magnetic field direction is given by
\begin{equation}
k = k_\parallel \cos^2 \theta + k_\perp \sin^2 \theta
\end{equation}
where $k_\perp$ is the diffusion coefficient perpendicular to the
magnetic field.  Jokipii (1987) shows that
\begin{equation}
k_\perp \approx {k_\parallel \over 1 + \eta^2}
\label{eq:k_perp_Jokipii87}
\end{equation}
provided that $\eta$ is not too large ($\eta$ values up to
$\sim$10), and that acceleration at perpendicular
shocks can be much faster than for the parallel case.  For
a perpendicular shock ($\theta=90^\circ$),
$k = k_\perp$ and $B_2 \approx 4 B_1$ and one obtains
\begin{equation}
t_{\rm acc}^\perp \approx {8 \over 3} {E \over \eta e B_1 u_1^2}.
\end{equation}
For a shock speed of $u_1 = 0.1 c$ and $\eta=10$ one obtains an
acceleration rate (in SI units) of
\begin{equation}
\left. {dE \over dt} \right|_{\rm acc} \approx 0.04 e c^2 B.
\end{equation}

Ellison, Baring and Jones (1995) have examined in detail the
acceleration time and injection efficiency for oblique shocks.
They point out that Eq.~\ref{eq:acc_rate} is only valid when
$k_1/u_1 \ge r_g$ which requires that cosmic ray velocities, $v$,
satisfy $\eta \le v/u_1$.  If this condition is not met, as would
typically be the case for thermal particles if $\eta \gg 1$, then
these particles would have a reduced probability of returning to
the shock after crossing.  As injection into the acceleration
process is assumed to be from the thermal plasma, having particle
speeds $v \sim u_1$, this will cause problems for the injection
of cosmic rays in the case of highly oblique shocks. Thus,
although having $\eta \gg 1$ can significantly reduce the
acceleration time in oblique shocks, the injection efficiency
would also be significantly reduced.  Ellison, Baring and Jones
(1995) find this to be a serious problem for $\theta > 40^\circ$.
One possible solution to this problem could be injection of a
previously accelerated particle population, e.g.\ acceleration in
a pulsar magnetosphere followed by injection into the supernova
shock.  Alternatively, injection from a supra-thermal tail of the
thermal distribution could help, with the supra-thermal tail
being due to heating by hard X-rays or gamma-rays from a nearby
source.

Supernova shocks remain strong enough to continue accelerating
cosmic rays for about 1000 years.  The rate at which cosmic rays
are accelerated is inversely proportional to the diffusion
coefficient (faster diffusion means less time near the shock).
For the maximum feasible acceleration rate, a typical
interstellar magnetic field, and 1000 years for acceleration,
energies of $10^{{14}} \times Z$ eV are in principle possible
($Z$ is atomic number) at parallel shocks, and $10^{{16}} \times
Z$ eV at perpendicular shocks but in this case, as noted above,
injection may be a problem.

%%%%%%%%%%%%%%%%%%%%%%%%%%%%%%%%%%%%%%%%%%%%%%%%%%%%%%%%%%%%%%%%%%%%%%%%%%%%%%%

\subsection{Effect of Cosmic Ray Pressure}

Inclusion of the effects of cosmic ray pressure on the shock
profile, and consequently on the spectrum of accelerated
particles, is a very difficult problem and we refer the reader to
Jones \& Ellison (1991) for a detailed discussion.  The shock
profile, instead of being a step-function becomes smoothed, and
this affects the acceleration of low and high energy particles
differently thereby affecting the cosmic ray spectrum.
Generally, the results are sensitive to Mach number, fraction of
energy flux of upstream plasma converted to accelerated
particles, energy dependence of diffusion coefficient, etc.
Calculated spectra may be $E^{-2}$, or flatter, or concave,
depending on the input and the approximations made.

The original method of treating this non-linear effect is the
two-fluid model (see e.g.\ the review by Drury 1983b), the two
fluids being plasma with ratio of specific heats $\gamma$=5/3,
and cosmic rays with $\gamma$=4/3.  To solve the steady-state
two-fluid equations it is necessary to make some approximations,
usually that the cosmic rays interact with the gas only through
their pressure, and that the cosmic ray pressure and energy flux
are continuous across the shock.  Furthermore an effective ratio
of specific heats $\gamma_{\rm eff}$ and energy-independent
diffusion was generally assumed.  The solutions were often found
to be unstable at high Mach numbers unless the spectrum was
cut-off artificially, and the approximations used meant that there was, in
effect, injection without conservation of particle
number.

In time-dependent two-fluid model calculations (e.g.\ Falle \&
Giddings 1987, Bell 1987), $\gamma_{\rm eff}$ can be calculated
self-consistently by weighting $\gamma$ by the pressures of the
two fluids taking account of the spectrum of energetic particles,
and energy-dependent diffusion can be included.  Also, finite
times for acceleration effectively eliminate the problem of
injection without conservation of particle number.  See Baring
(1997) for a brief review and additional references, and
Blandford and Eichler (1987), Berezhko (1999) and Berezhko and
V\"olk (2000) for alternative approaches to this non-linear
problem.

%%%%%%%%%%%%%%%%%%%%%%%%%%%%%%%%%%%%%%%%%%%%%%%%%%%%%%%%%%%%%%%%%%%%%%%%%%%%%%%

\subsection{Relativistic Shocks}

Shocks in jets of active galactic nuclei (AGN) and GRB are likely
to be relativistic, i.e. $u_1 > 0.1c$, with shock Lorentz
factors, $\Gamma \sim 10$ and $\Gamma \sim 300$, respectively.
For relativistic plasma motion with bulk velocities comparable to
those of the particles being accelerated, the particle
distribution will not be isotropic, and the approximations used
earlier are no longer valid.  Instead, the typical escape
probability and fractional energy gain per shock crossing are
(very crudely) $\sim 0.5$ and $\sim 1$, respectively.  Initial
work on relativistic shock acceleration was done by Peacock
(1981); see Kirk \& Duffy (1999)
for a topical review and additional
references.

\begin{figure}[htb]
\epsfig{file=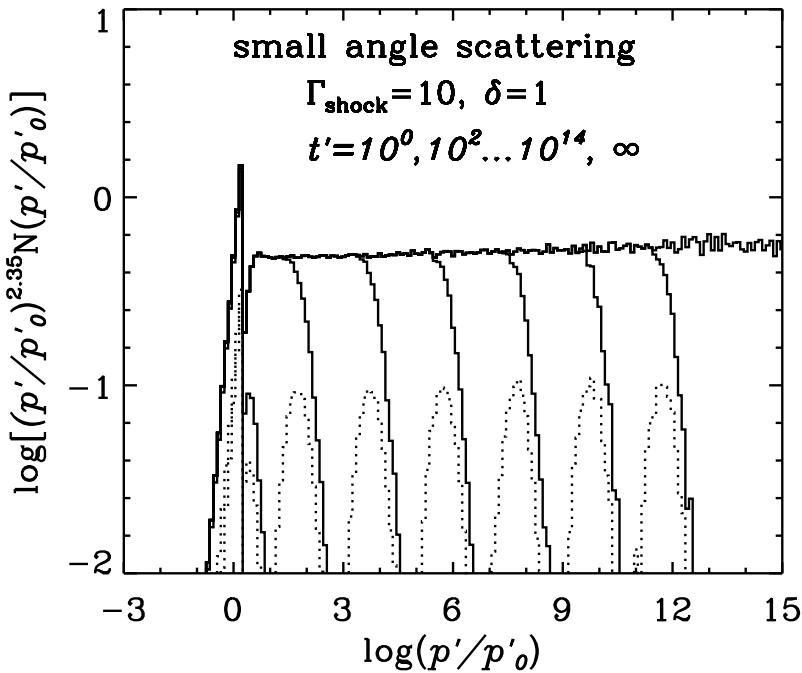,height=10cm}
\caption{Momentum spectra obtained from Monte Carlo simulations
with small angle scattering upstream for $\Gamma_1=10$ and $k
\propto p$ ($\delta=1$).  Spectra of particles which have escaped at times
$t'/t_0^{\rm sc}=10^0,10^1,\dots,10^{14}$ after injection are
shown (solid histograms) together with the spectra of particles
remaining in the acceleration zone (dotted histograms) at these
times (adapted from Protheroe 2001).
\label{fig:RSAprotheroe}}
\end{figure}

The techniques used have been analytic (eigenfunction method) and
Monte Carlo methods to solve the steady state equations for the
particle spectral and angular distributions.  The acceleration
time depends strongly on $u_1$, $k_\perp/k_\parallel$, and
$\theta$ and can be as low as $\sim (1$---10$)r_g/c$
(e.g. Bednarz \& Ostrowski~1996, 1998, and
Ostrowski~1998).
Detailed studies have shown a trend in which shocks with larger
$u_1$ generally have lower spectral indices (Kirk and Schneider 1987,
Ellison and Jones 1990), and
the spectral index can be very sensitive to the pitch angle
particularly for mildly-relativistic shocks (see Baring
1997 for additional references).  Nevertheless,
for plane ultra-relativistic shocks, test-particle Monte Carlo
simulations tend to give spectral indices close to $\Gamma =
2.25$ (e.g. Bednarz \& Ostrowski~1998, Gallant, Achterberg \&
Kirk 1999, Bednarz 2000, Kirk et al.\ 2000, Achterberg et al.\
2001, Protheroe 2001, Meli and Quenby 2003a,2003b).
These simulations assume strong turbulence downstream of the
shock, but if this not present then the spectral index will be
steeper (Ostrowski \& Bednarz 2002).

Shock modification by the back-reaction of accelerated particles
(Ellison \& Double 2002) can cause the
compression ratio to increase above the test particle value
causing the spectrum of accelerated particles to differ from a
simple power law with $\Gamma \approx 2.25$ for shock Lorentz
factors less than 10.  

A recent development has been to simulate separately the
propagation upstream and downstream to work out the probability
of returning to the shock at a particular angle to the shock
normal for a given direction on crossing the shock (Protheroe
2001, Lemoine \& Pelletier 2003), and use these
distributions to simulate very efficiently relativistic shock
acceleration over a large dynamic range in particle momentum.
The time evolution of the momentum spectrum for injection at the
shock, at time $t=0$, of highly relativistic mono-energetic
particles which are isotropic in the upstream frame and have
upstream-frame momentum $p_0$ as calculated by Protheroe
(2001) is shown in
Fig.~\ref{fig:RSAprotheroe}.  Here, as observed in the downstream
frame these injected particles have a range of initial momenta
distributed up to $\sim 2 p_0'$ where $p_0'=\Gamma_{12}p_0$ 
($\Gamma_{12}$ is the Lorentz factor for transforming between
upstream and downstream frames).  All distances are measured in
units of the upstream scattering mean free path $\lambda^{\rm
sc}_0$ and all times are measured in units of $t^{\rm
sc}_0=\lambda^{\rm sc}_0/c$.  The figure shows how the spectrum
develops with time after injection, showing separately at each
time indicated the spectrum of particles which have escaped, and
(dotted curves) the spectrum of particles remaining within the
acceleration zone (i.e.\ not having yet escaped downstream).

%%%%%%%%%%%%%%%%%%%%%%%%%%%%%%%%%%%%%%%%%%%%%%%%%%%%%%%%%%%%%%%%%%%%%%%%%%%%%%%

\section{Interactions of High Energy Cosmic Rays}

Interactions of cosmic rays with radiation are important both
during acceleration when the resulting energy losses compete with
energy gains by, for example, shock acceleration, and during
propagation from the acceleration region to the observer.  For
UHE CR the most important processes are pion photoproduction and
Bethe-Heitler pair production both on the CMBR, and synchrotron
radiation.  In the case of nuclei, photodisintegration on the
CMBR is also important.

\subsection{Nucleons}

The mean interaction length, $x_{p \gamma}$, of a proton of
energy $E$ is given by,
\begin{equation}
        {1 \over x_{p \gamma}(E)}= {1 \over 8 \beta E^2}
\int_{\varepsilon_{\rm min}(E)}^
        {\infty} \frac{n(\varepsilon)}{\varepsilon^2} 
	\int_{s_{\rm min}}^{s_{\rm max}(\varepsilon,E)} \hspace{-3mm}
        \sigma(s)(s-m_p^2 c^4)ds d\varepsilon,
        \label{eq:mpl}
\end{equation}
where $n(\varepsilon)$ is the differential photon number density
of photons of energy $\varepsilon$, and $\sigma(s)$ is the
appropriate total cross section for the process in question for a
centre of momentum (CM) frame energy squared, $s$, given by
\begin{equation}
s=m_p^2 c^4 + 2 \varepsilon E(1 - \beta \cos \theta)
\label{eq:s}
\end{equation}
where $\theta$ is the angle between the directions of the proton
and photon, and $\beta c$ is the proton's velocity.  For pion
photoproduction $s_{\rm min} \approx 1.16$ GeV$^2$, and for
Bethe-Heitler pair production the threshold is somewhat lower,
$s_{\rm min} \approx 0.882$ GeV$^2$.  For both processes,
$s_{\rm max} \approx (m_p^2c^4+4\varepsilon E)$ which corresponds
to a head-on collision of a proton of energy $E$ and a photon of
energy $\varepsilon$.

\begin{figure}[htb]
{\hspace*{-5em}\epsfig{file=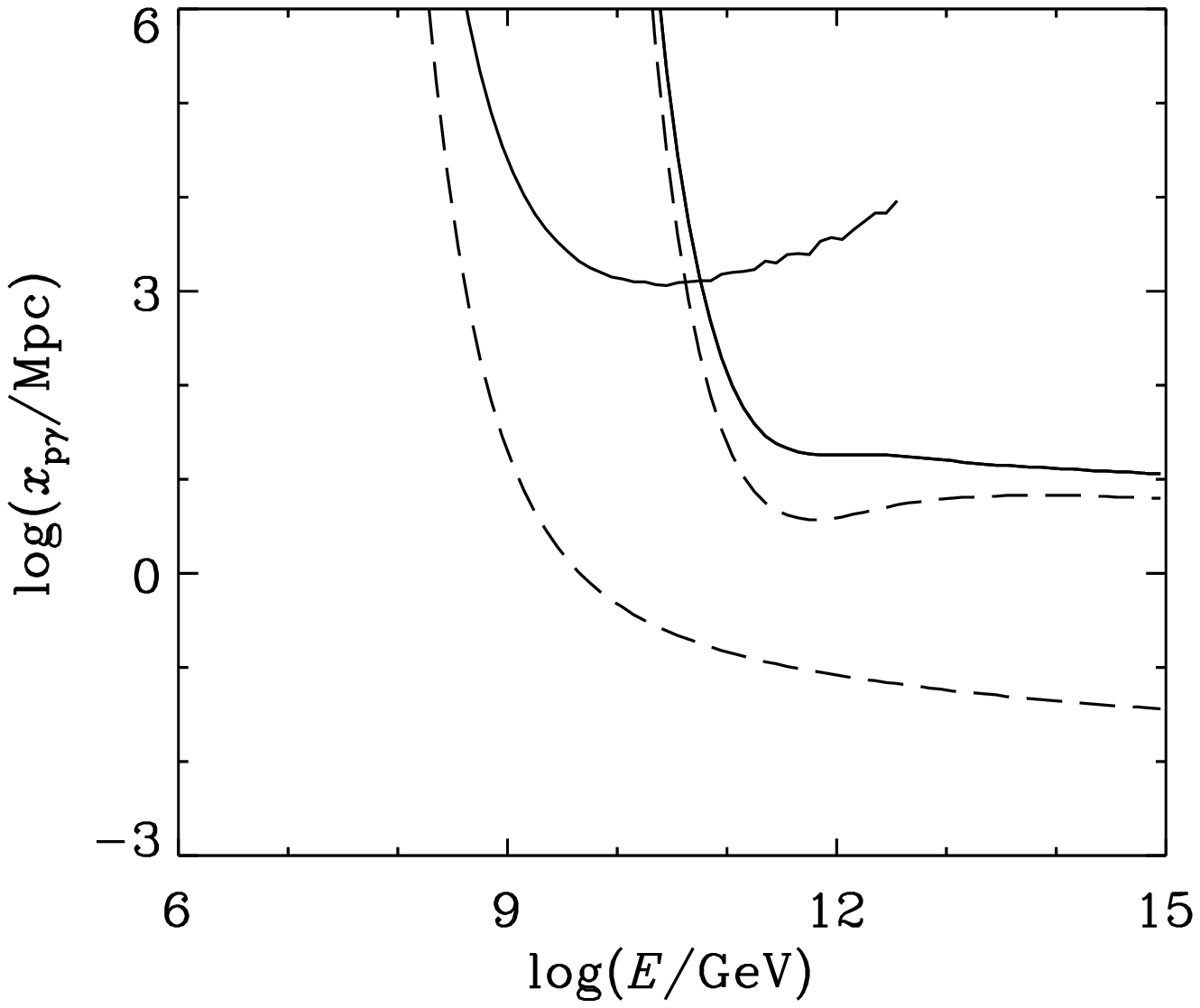,height=8cm}\hspace*{-6em}\epsfig{file=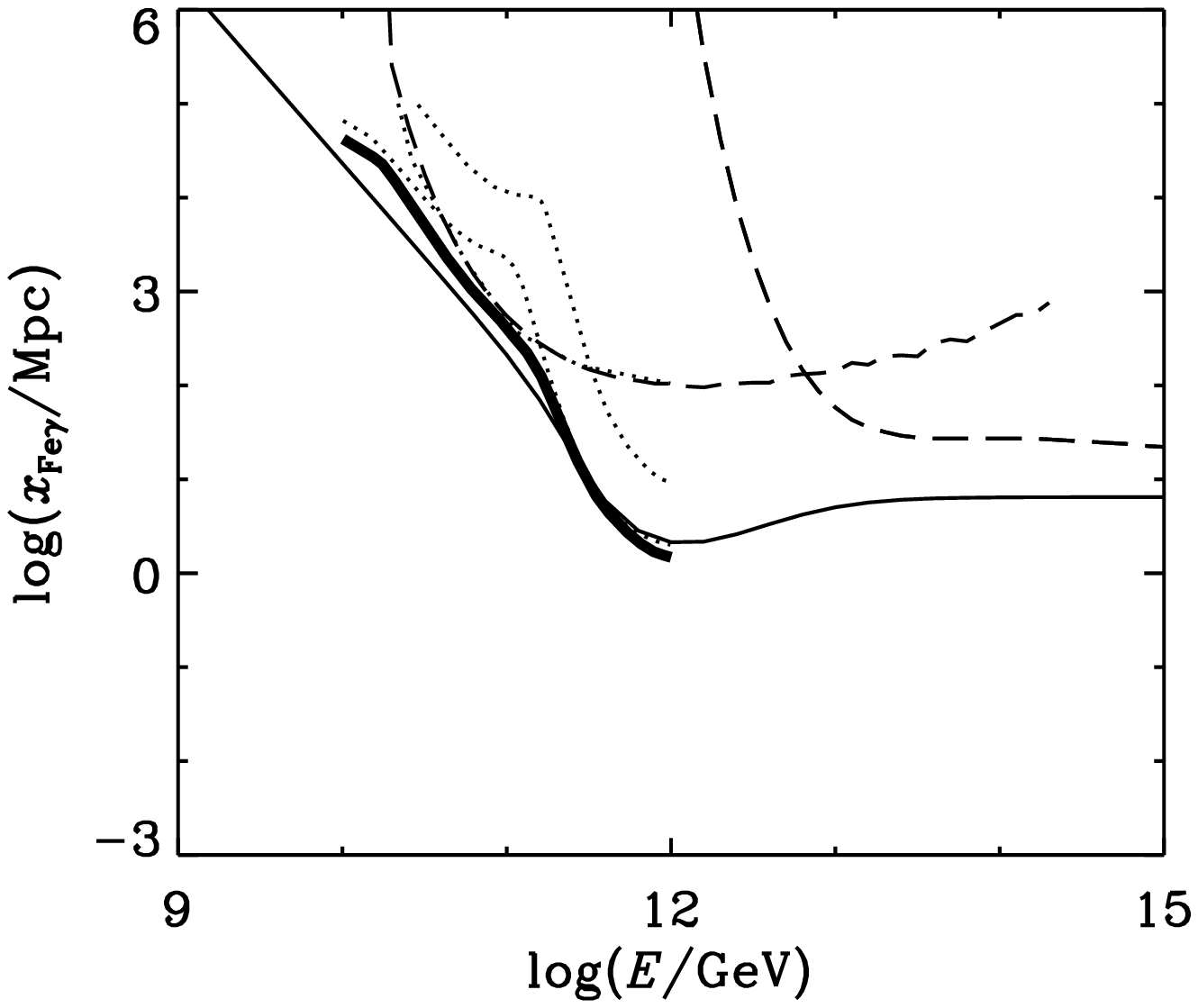,height=8cm}}
\caption{(a) Mean interaction length (dashed lines) and
energy-loss distance (solid lines), $E/(dE/dx)$, for
Bethe-Heitler pair production and pion photoproduction in the
CMBR (lower and higher energy curves respectively).  (From
Protheroe and Johnson 1995).
(b) Energy-loss distance of Fe-nuclei in the CMBR for
pair-production (leftmost dashed line) and pion photoproduction
(rightmost dashed line).  Photodisintegration distances are given
for loss of one nucleon (lower dotted curve), two nucleons (upper
dotted line) as well as the total loss distance (thick curve) estimated
by Stecker and Salamon (1999).  The thin full curve shows an
estimate over a larger range of energy (Protheroe, unpublished)
of the total loss distance based on photodisintegration cross
sections of Karakula and Tkaczyk (1993).
\label{fig:pgpiee3k_xloss}}
\end{figure}

The mean interaction lengths for both processes,
$x_{p\gamma}(E)$, are obtained from Equation \ref{eq:mpl} for
interactions in the CMBR and are plotted as dashed lines in
Fig. \ref{fig:pgpiee3k_xloss}(a).  Dividing by the mean
inelasticity of the collision, $\kappa(E)$, one obtains the
energy-loss distances for the two processes (solid curves),
\begin{equation}
{E \over dE/dx} = {x_{p \gamma}(E) \over \kappa(E)}.
\end{equation}

%%%%%%%%%%%%%%%%%%%%%%%%%%%%%%%%%%%%%%%%%%%%%%%%%%%%%%%%%%%%%%%%%%%%%%%%%%%%%%%
 
\subsection{Nuclei}

In the case of nuclei the situation is a little more complicated.
The threshold condition for Bethe-Heitler pair production can be
expressed as
\begin{equation}
\gamma > {m_e c^2 \over \varepsilon} \left( 1 + {m_e \over Am_p} \right),
\end{equation}
and the threshold condition for pion photoproduction can be
expressed as
\begin{equation}
\gamma > {m_\pi c^2 \over 2 \varepsilon} \left( {1 + {m_\pi \over
2A m_p}} \right).
\end{equation}
Since $\gamma = E / A m_p c^2$, where $A$ is the mass number, we
will need to shift both energy-loss distance curves in
Fig.~\ref{fig:pgpiee3k_xloss}(a) to higher energies by a factor of
$A$.  We shall also need to shift the curves up or down as
discussed below.

For Bethe-Heitler pair production the energy lost by a nucleus in
each collision near threshold is approximately $\Delta E \approx
\gamma 2 m_e c^2$.  Hence the inelasticity is
\begin{equation}
\kappa \equiv { \Delta E \over E} \approx {2m_e \over A m_p},
\end{equation}
and is a factor of $A$ lower than for protons.  On the other
hand, the cross section goes like $Z^2$, so the overall shift is
down (to lower energy-loss distance) by $Z^2/A$.  For example,
for iron nuclei the energy loss distance for pair production is
reduced by a factor $26^2/56 \approx 12.1$.

For pion production the energy lost by a nucleus in each
collision near threshold is approximately $\Delta E \approx
\gamma m_\pi c^2$, and so, as for pair production, the
inelasticity is factor $A$ lower than for protons.  The cross
section increases approximately as $A^{0.9}$ giving an overall
increase in the energy loss distance for pion production of a
factor $\sim A^{0.1} \approx 1.5$ for iron nuclei.
The energy loss distances for pair production and pion
photoproduction are shown for iron nuclei in
Fig.~\ref{fig:pgpiee3k_xloss}(b) 

Photodisintegration can be very important both during
acceleration and propagation and has been considered in detail by
Tkaczyk et al. (1975), Puget et al. (1976), Karakula and Tkaczyk
(1993), Epele and Roulet (1998) and Stecker and Salamon (1999).
The photodisintegration distance defined by $A/(dA/dx)$ taken
from Stecker and Salamon (1999) is shown in
Fig.~\ref{fig:pgpiee3k_xloss}(b) together with an estimate made
over a larger range of energy by Protheroe (unpublished) of the
total loss distance based on photodisintegration cross sections
of Karakula and Tkaczyk (1993).  Clearly, photodisintegration is
the dominant loss process for iron nuclei.

%%%%%%%%%%%%%%%%%%%%%%%%%%%%%%%%%%%%%%%%%%%%%%%%%%%%%%%%%%%%%%%%%%%%%%%%%%%%%%%

\section{Maximum Energies}

Because of their much lower energy losses at a given energy,
protons and nuclei can be accelerated to much higher energies
than electrons for a given magnetic environment.  For stochastic
particle acceleration by electric fields induced by motion of
magnetic fields $B$, the rate of energy gain by relativistic
particles of charge $Ze$ can be written (in SI units) $dE/dt =
\xi Ze c^2 B$ as in Eq.~1, where $\xi < 1$ and depends on the
acceleration mechanism; a value of $\xi =0.04$ might be achieved
by first order Fermi acceleration at a perpendicular shock with
shock speed $\sim 0.1 c$.

\begin{figure}[htb]
\hspace*{-5em}\epsfig{file=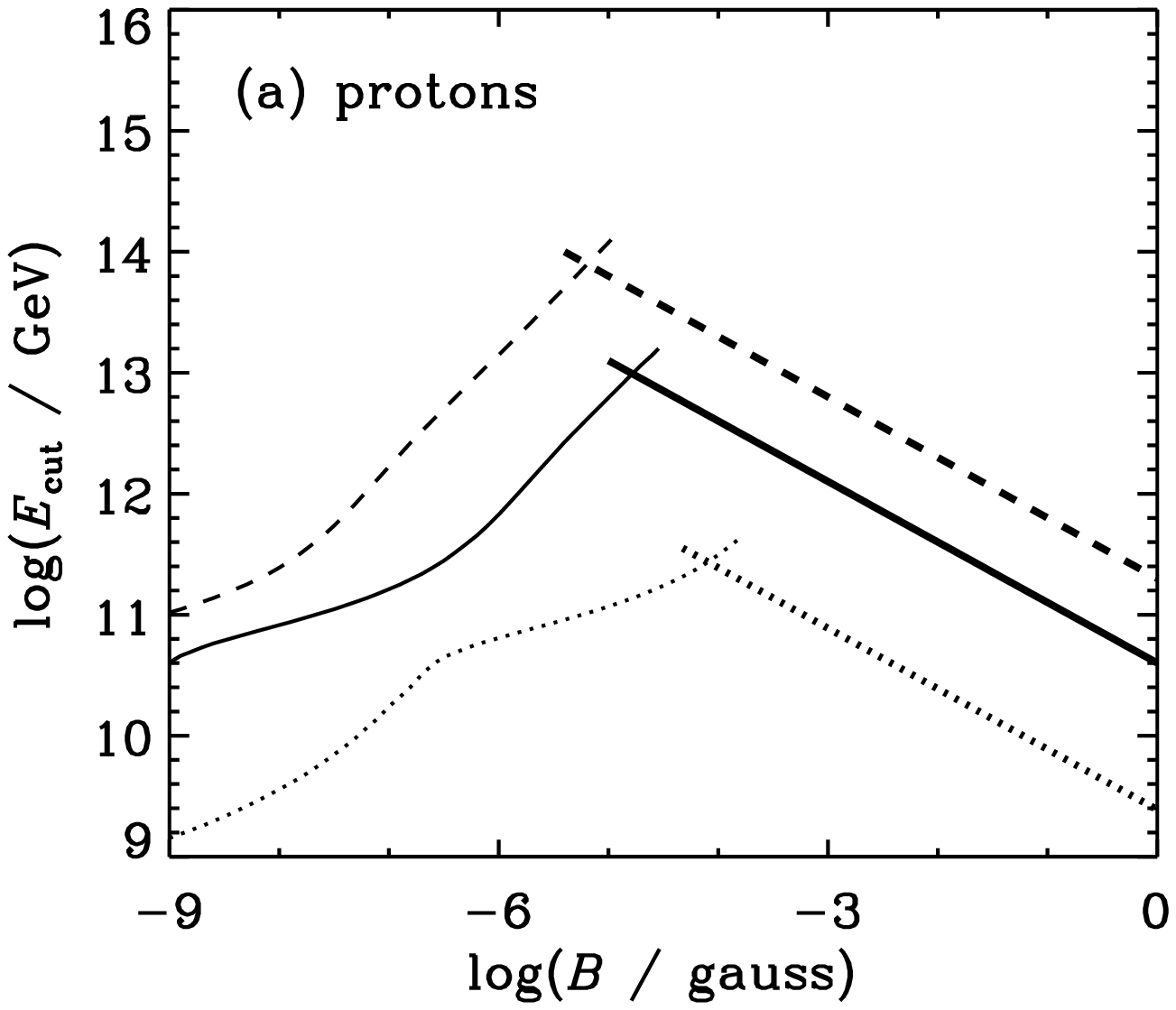,height=8cm}\hspace*{-6em}\epsfig{file=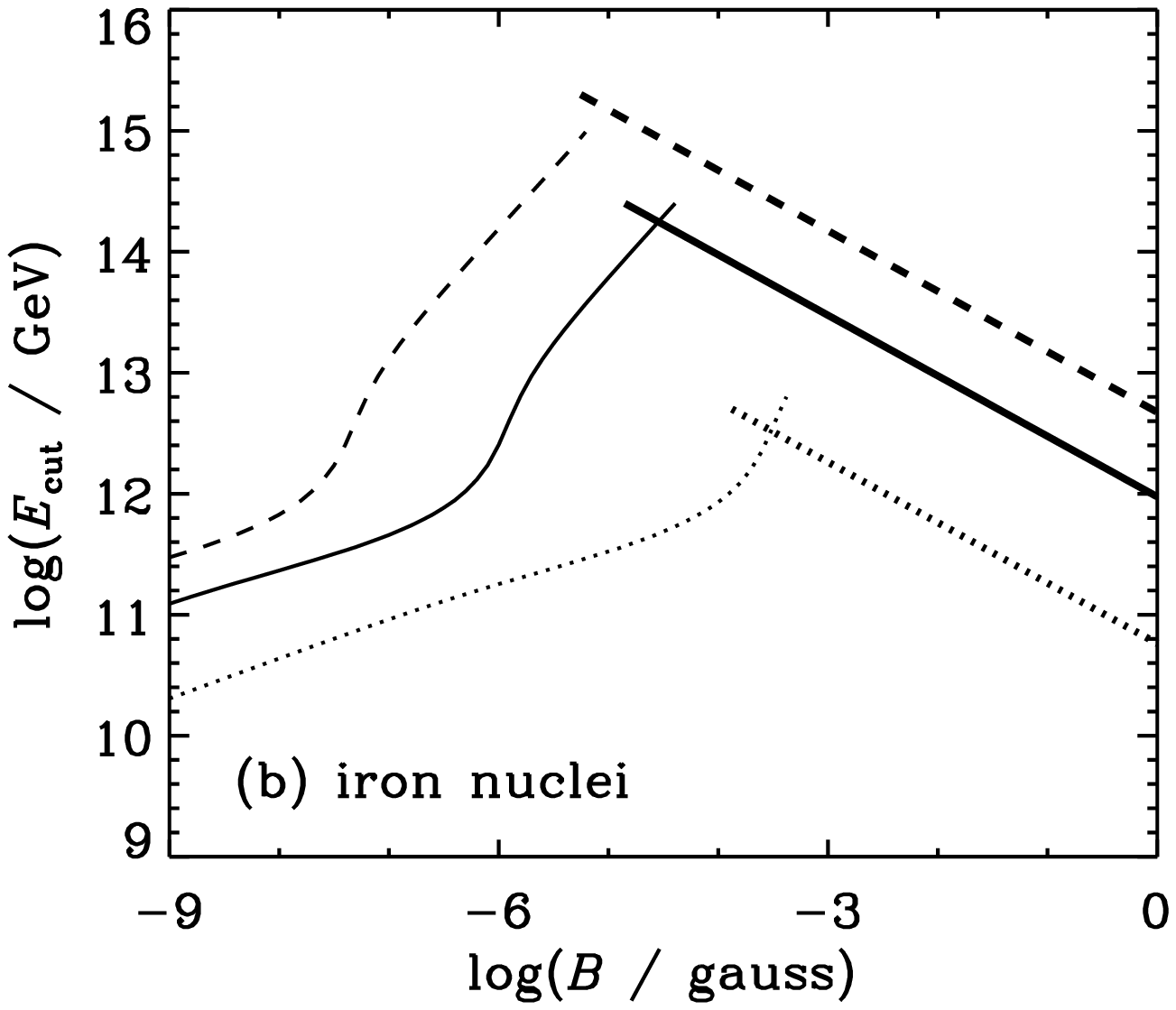,height=8cm}
\caption{Maximum energy as a function of magnetic field of (a)
protons and (b) iron nuclei for maximum possible acceleration
rate $\xi=1$ (dashed), plausible acceleration at perpendicular
shock $\xi=0.04$ (solid), and plausible acceleration at parallel
shock $\xi=1.5 \times 10^{-4}$ (dotted).  Straight lines on the
right give the limit from synchrotron loss, curved lines on the
left give the limit from Bethe-Heitler pair production and pion
photoproduction (protons) or photodisintegration (iron nuclei).
\label{fig:pgpiee3k_emax}}
\end{figure}

The rate of energy loss by synchrotron radiation of a particle of mass
$Am_p$, charge $Ze$, and energy $\gamma m c^2$ is
\begin{equation}
- \left. {dE \over dt} \right|_{\rm syn} = {4 \over 3} \sigma_T
\left( {Z^2 m_e \over Am_p} \right)^2 {B^2 \over 2 \mu_0} \gamma^2 c.
\end{equation}
Equating the rate of energy gain with the rate of energy loss by
synchrotron radiation places one limit on the maximum energy achievable
by electrons, protons and nuclei:
\begin{eqnarray}
E_e^{\rm cut} &=& 6.0 \times 10^{2} \xi^{1/2}
\left( {B \over {\rm 1 \; T}} \right)^{-1/2} \hspace{5mm} {\rm GeV}, \\
E_p^{\rm cut} &=& 2.0 \times 10^{9} \xi^{1/2}
\left( {B \over {\rm 1 \; T}} \right)^{-1/2} \hspace{5mm} {\rm GeV}, \\
E_{Z,A}^{\rm cut} &=& 2.0 \times 10^{9} \xi^{1/2} {A^2 \over Z^{3/2}}
\left( {B \over {\rm 1 \; T}} \right)^{-1/2} \, {\rm GeV}.
\end{eqnarray}
The cut-off energies of protons and iron nuclei allowed by
synchrotron radiation losses are shown in
Figs.~\ref{fig:pgpiee3k_emax}(a) and \ref{fig:pgpiee3k_emax}(b) respectively,
and are plotted against magnetic field for three values of $\xi$.

Equating the total energy loss rate for proton--photon
interactions (i.e. the sum of pion production and Bethe-Heitler
pair production) in Fig.~\ref{fig:pgpiee3k_xloss}(a) to the rate
of energy gain by acceleration gives the maximum proton energy in
the absence of other loss processes.  This is shown in
Fig.~\ref{fig:pgpiee3k_emax}(a) for the three $\xi$ values.  As
can be seen, for a perpendicular shock it is possible in
principle to accelerate protons to $\sim 10^{13}$ GeV in a $\sim
10^{-5}$ G field.

The effective loss distance given in
Fig.~\ref{fig:pgpiee3k_xloss}(b) is used together with the
acceleration rate for iron nuclei to obtain the maximum energy as
a function of magnetic field.  This is shown in
Fig.~\ref{fig:pgpiee3k_emax}(b).  We see that for a perpendicular
shock it is in principle possible to accelerate iron nuclei to
$\sim 2 \times 10^{14}$ GeV in a $\sim 3 \times 10^{-5}$ G field.
While this is higher than for protons, iron nuclei are likely to
get photodisintegrated into nucleons of maximum energy $\sim 4
\times 10^{12}$ GeV, and so there is not much to be gained unless
the source is nearby.  

Of course, potential acceleration sites need to have the
appropriate combination of size (much larger than the gyroradius
at the maximum energy), magnetic field, shock velocity (or other
relevant velocity) and the time available for acceleration.
These limits were obtained and discussed in some detail by
Biermann and Strittmatter (1987).  We emphasize that
the cut-off energies estimated above apply to ideal conditions in
which strong scattering occurs at all energies between injection
and the cut-off energy; in reality, the $\xi$-values above are probably
optimistic.

%%%%%%%%%%%%%%%%%%%%%%%%%%%%%%%%%%%%%%%%%%%%%%%%%%%%%%%%%%%%%%%%%%%%%%%%%%%%%%%

\section{Spectral Shape near Maximum Energy}

To determine the spectral shape near maximum energy we use the
leaky-box acceleration model (Szabo and Protheroe 1994, Protheroe
and Stanev 1998) which
may be considered as follows.  A particle of energy $E_0$ is
injected into the leaky box.  While inside the box, the
particle's energy changes at a rate $dE/dt = E r_{\rm acc}(E)$
and in any short time interval $\Delta t$ the particle has a
probability of escaping from the box given by $\Delta t r_{\rm
esc}(E)$.  The energy spectrum of particles escaping from the box
then approximates the spectrum of shock accelerated particles.

At time $t$ after injecting $N_0$ particles of momentum $E_0$
into the acceleration zone, the number of particles remaining in
the acceleration zone, $N_{\rm rem}(t)$, is obtained by solving
\[
{dN_{\rm rem} \over dt} = -N_{\rm rem}(t) r_{\rm esc}[E(t)]
\]
which has solution
\[
N_{\rm rem}(t)=N_0 \exp \left[ - \int_0^t r_{\rm esc}[E(t)] dt \right] = 
N_0 \exp \left[ - \int_{E_0}^E r_{\rm esc}(E) {dt \over dE}dE\right] .
\]
The spectrum of accelerated (escaping) particles is then 
\begin{eqnarray}
{dN \over dE} &=&  -{dN_{\rm rem} \over dt} {dt \over dE}
= {N_{\rm rem}[t(E)]r_{\rm esc}[E(t)] \over E \, r_{\rm acc}[E(t)]} \nonumber  \\ 
{dN \over dE} &=& N_0 {r_{\rm esc}(E) \over E \, r_{\rm acc}(E)}
\exp \left[ - \int_{E_0}^E { r_{\rm esc}(E)  \over E \, r_{\rm acc}(E)}dE\right] .
\label{eq:fund}
\end{eqnarray}

Let us consider first the case of no energy losses, interactions,
or losses due to any other process.  Assuming that the diffusion
coefficients upstream and downstream have the same power-law
dependence on energy, and using Eq.~\ref{eq:esc_acc},
\begin{equation}
r_{\rm acc} = a E^{-\delta}, \;\;\; r_{\rm esc} = (\Gamma - 1)a E^{-\delta}.
\end{equation}
Then the differential energy spectrum of particles which
have escaped from the accelerator is given by
\begin{eqnarray}
dN/dE & = & N_0 (\Gamma - 1) (E_0)^{- 1} (E/E_0)^{- \Gamma}, 
\label{eq:spec_nocut}
\end{eqnarray}
for $(E>E_0)$ where $\Gamma = (R+2)/(R-1)$ is the differential spectral index.  

\subsection{Cut-Off due to Finite Acceleration Volume, etc.}

Even in the absence of energy losses, acceleration usually ceases
at some energy due to the finite size of the acceleration volume
(e.g. when the gyroradius becomes comparable to the characteristic
size of the shock), or as a result of some other process.  We
approximate the effect of this by introducing a constant term to
the expression for the escape rate:
\begin{equation}
r_{\rm esc} = (\Gamma - 1)a E^{-\delta} + (\Gamma - 1)a E_{\rm max}^{-\delta}.
\label{eq:escrate}
\end{equation}
where $E_{\rm max}$ is defined by the above equation and will
be close to the energy at which the spectrum steepens due
to the constant escape term.  We shall refer to $E_{\rm max}$ 
as the ``maximum energy'' even though some particles will be
accelerated to energies above this.

\begin{figure}[htb]
\hspace*{-5em}\epsfig{file=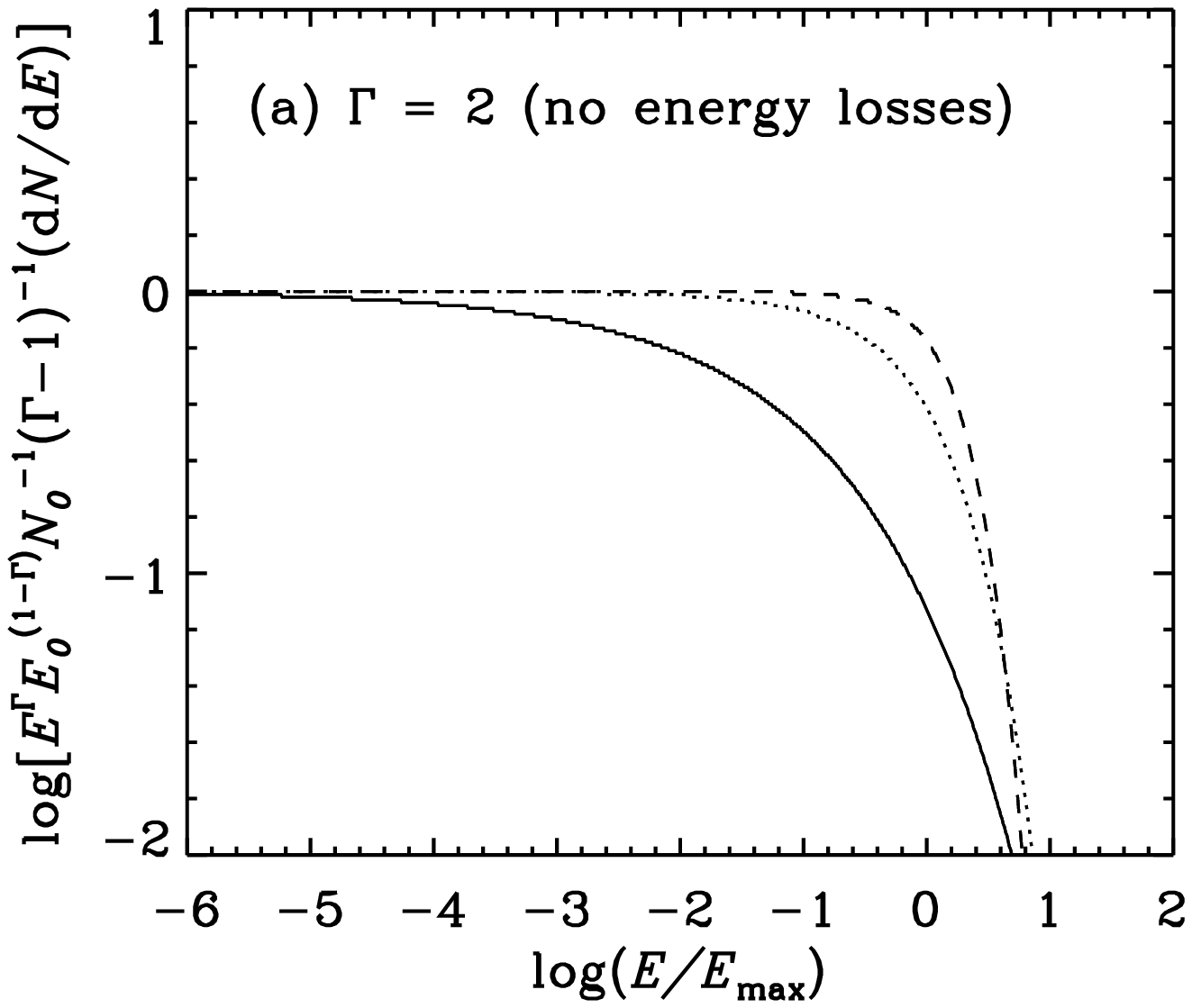,height=8cm}\hspace*{-6em}\epsfig{file=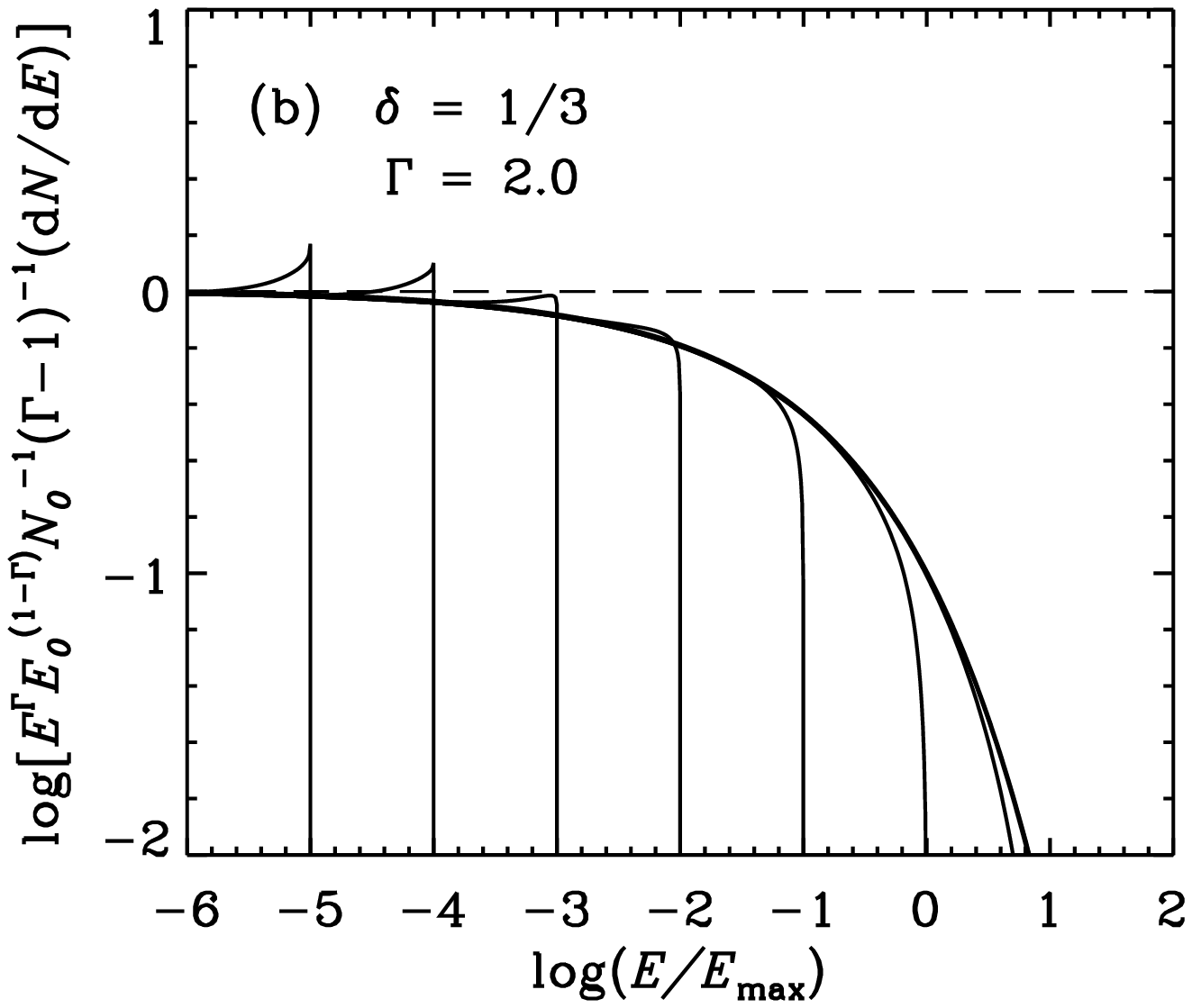,height=8cm}
\caption{(a) Differential energy spectrum for the case of a cut-off
due to escape for $\Gamma=2$ and $\delta=1/3$ (solid curve), 2/3
(dotted curve) and 1 (dashed curve) -- from Protheroe \& Stanev
(1998).  (b) Differential energy spectrum for the case of a cut-off
due to $E^2$ energy losses for $\Gamma=2$ and $\delta=1/3$ and
$E_{\rm cut}/E_{\rm max} = 10^{-5}$ (leftmost curve), $10^{-4}$,
\dots , $10^{2}$ (rightmost curve).}
\label{leakage_cutoff_comp_synch}
\end{figure}

Following the same procedure as for the case of a purely
power-law dependence of the escape rate, one  obtains the
differential energy spectrum of particles $(E>E_0)$ escaping from the
accelerator,
\begin{eqnarray}
{dN \over dE} \! &=& \! N_0 (\Gamma - 1) (E_0)^{- 1} (E/E_0)^{-
\Gamma}[ 1 + ( {E / E_{\rm max}} )^\delta ] \nonumber \\ && \hspace{-5mm}
\times \exp \left\{ - {\Gamma - 1 \over \delta} \left[ \left( {E
\over E_{\rm max}} \right)^\delta - \left( {E_0 \over E_{\rm
max}} \right)^\delta \right] \right\}
\end{eqnarray}
for $\delta>0$ (Protheroe and Stanev
1998).  We compare in
Fig~\ref{leakage_cutoff_comp_synch}(a) the spectra for $\Gamma=2$
and $\delta$ ranging from $1/3$ to $1$, and note that the energy
dependence of the diffusion coefficient has a profound influence
on the shape of the cut-off.  This smooth cut-off occurs over up
to three decades in energy and its shape depends on the momentum
dependence of the diffusion coefficients, and the intrinsic
spectral index $\Gamma$ which depends on the
compression ratio.  Such a smooth cut-off has very recently been
noted in test-particle Monte Carlo simulations of shock
acceleration at shocks in a cylindrical jet geometry where there
is sideways leakage out of the jet (Casse and Markowith,
2003).

\subsection{Cut-Off due to Energy Losses}

When continuous energy losses are included the spectrum is
cut-off sharply at an energy at which the total rate of energy
gain is zero.  Depending on the spectral index, and momentum
dependence of the diffusion coefficient, either a pile-up or a
steepening in the spectrum occurs before the cut-off.  To
calculate the energy spectrum a term representing the energy-loss
rate must be added to the acceleration rate,
\begin{equation}
r_{\rm acc} = a E^{-\delta}+ \left. {d E \over dt}\right|_{\rm loss},
\end{equation}
but since the physical size of the ``box'' increases with energy,
synchrotron losses can cause a particle in the downstream region
to effectively fall out of the box (Drury et al 1999).  This
process can be represented by an additional escape term in the
escape rate.  The acceleration zone extends distances
$L_1(E)=k_1(E)/u_1$ and $L_2(E)=k_2(E)/u_2$ upstream and
downstream of the shock.  With the additional term in the rate of
escape of particles due to energy loss,
\begin{eqnarray}
r_{\rm esc} &=& (\Gamma-1)a E^{-\delta} + (\Gamma-1)a E_{\rm
max}^{-\delta} + {1 \over L_1(E)+L_2(E)} {d L_2 \over dE}\left(-
\left. \;{d E \over dt}\right|_{\rm loss}\right)\\ r_{\rm esc}
&=& (\Gamma-1)a (E^{-\delta} + E_{\rm max}^{-\delta}) +
\delta{L_2 \over L_1+L_2} r_{\rm loss}(E)
\label{eq:escrate_with_Eloss}
\end{eqnarray}
where $r_{\rm loss}(E)=-(dE/dt)_{\rm loss}/E$, and we adopt
$L_2/(L_1+L_2)=R/(1+R)$ following Drury et al.\ (1999).

For the case of synchrotron losses the result depends on the
parameters $\delta$, $\Gamma$, $E_0$, $E_{\rm cut}$ and $E_{\rm
max}$.  As a result of the energy loss by particles near the
cut-off energy, a pile-up in the spectrum may be produced just
below $E_{\rm cut}$.  The size of the pile-up will be determined
by the relative importance of $r_{\rm acc}$ and $r_{\rm esc}$ at
energies just below $E_{\rm cut}$.  Numerical solution of
Eq.~\ref{eq:fund} gives the results for $\Gamma=2$, $\delta=1/3$
and various $E_{\rm max}$ which are shown in
Fig.~\ref{leakage_cutoff_comp_synch}(b).

\begin{figure}[htb]
\epsfig{file=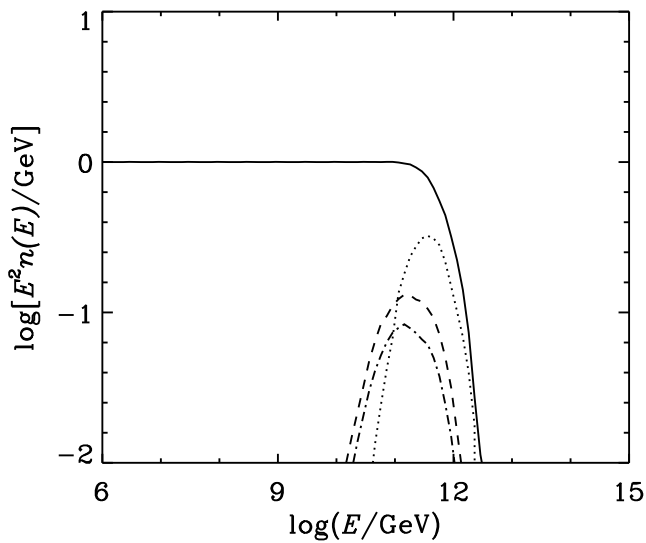,height=8cm}
\caption{The spectrum of particles produced during acceleration
(multiplied by $E^2$) per proton injected into the accelerator
($\Gamma=2$, $k \propto E$): protons (full curve), neutrons
(dotted curve), charged pions (dashed curve) and neutral pions
(chain curve). Results are shown for $E_{\rm cut}= 2 \times
10^{12}$ GeV due to photoproduction on the cosmic microwave
background.  (Adapted from figure~7 of Szabo and Protheroe 1994).
\label{fig:AccSpec98}}
\end{figure}

One can use the Monte Carlo method to investigate the shape of
the cut-off or pile-up which results when the nominal cut-off
energy is determined by interactions rather than continuous
energy losses.  This technique was used by Protheroe and Stanev
(1998) to investigate cut-offs in electron spectra due to inverse
Compton scattering in the Klein-Nishina regime, and by Szabo and
Protheroe(1994) to investigate cut-offs in the proton spectrum
due to photoproduction in a radiation field.  Results for $E_{\rm
cut}= 2 \times 10^{12}$ GeV due to photoproduction on the CMBR
are shown in Figure~\ref{fig:AccSpec98}.

%%%%%%%%%%%%%%%%%%%%%%%%%%%%%%%%%%%%%%%%%%%%%%%%%%%%%%%%%%%%%%%%%%%%%%%%%%%%%%%
 
\section{Cascading in Cosmic Radiation Fields}

As well as particles being produced during the acceleration
process as a result of interactions, during propagation to Earth
cascading occurs and the accompanying fluxes of $\gamma$-rays and
neutrinos must not exceed the observed flux or flux limits.  By
measuring the accompanying fluxes, we may well provide additional
clues to the nature and origin of the highest energy cosmic rays
(Waxman and Bahcall 1999, Mannheim et al. 2001).  Hence it is
important to calculate these fluxes resulting from cascading.

\begin{figure}[htb]
\hspace*{-5em}\epsfig{file=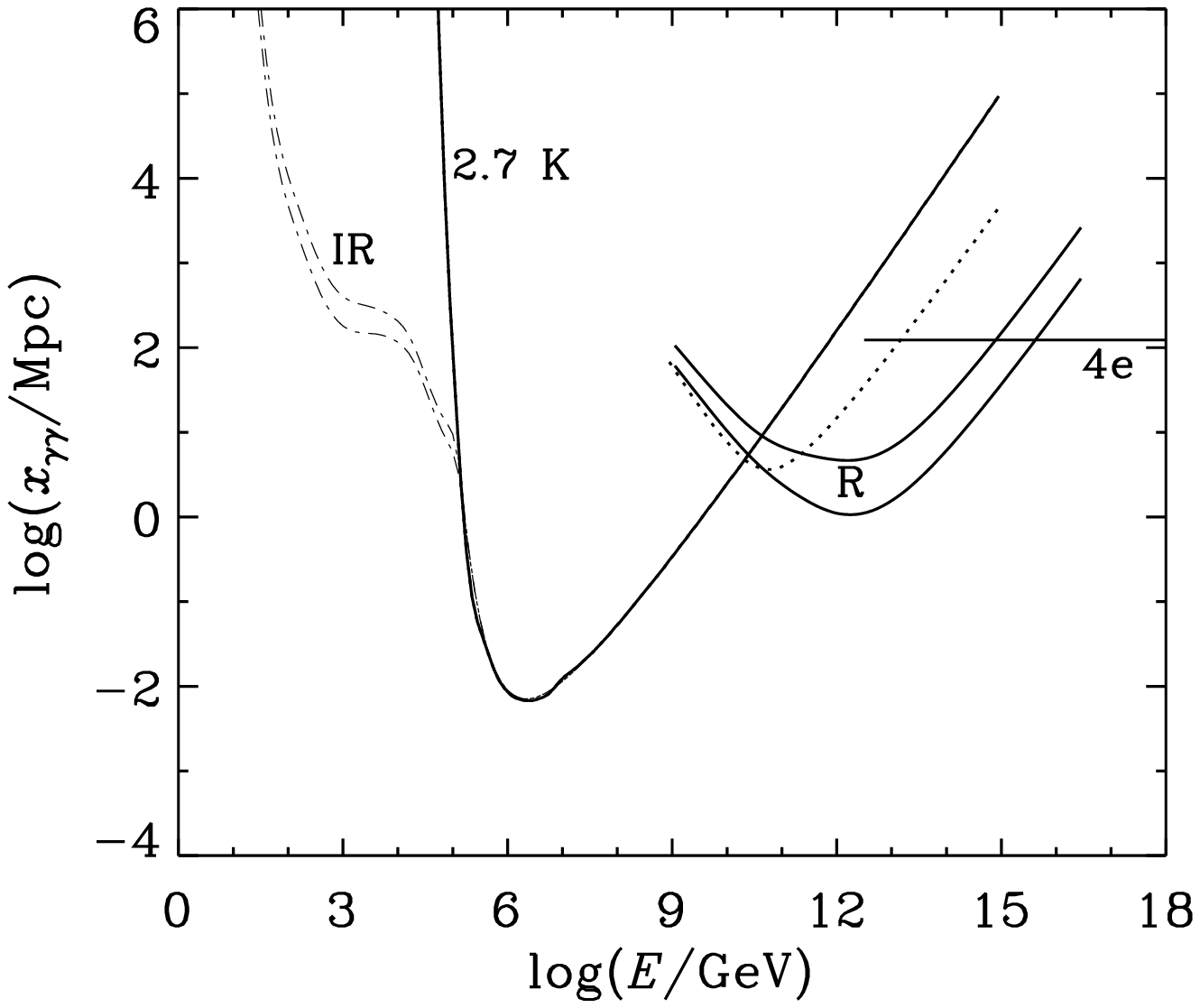,height=8cm}\hspace*{-6em}\epsfig{file=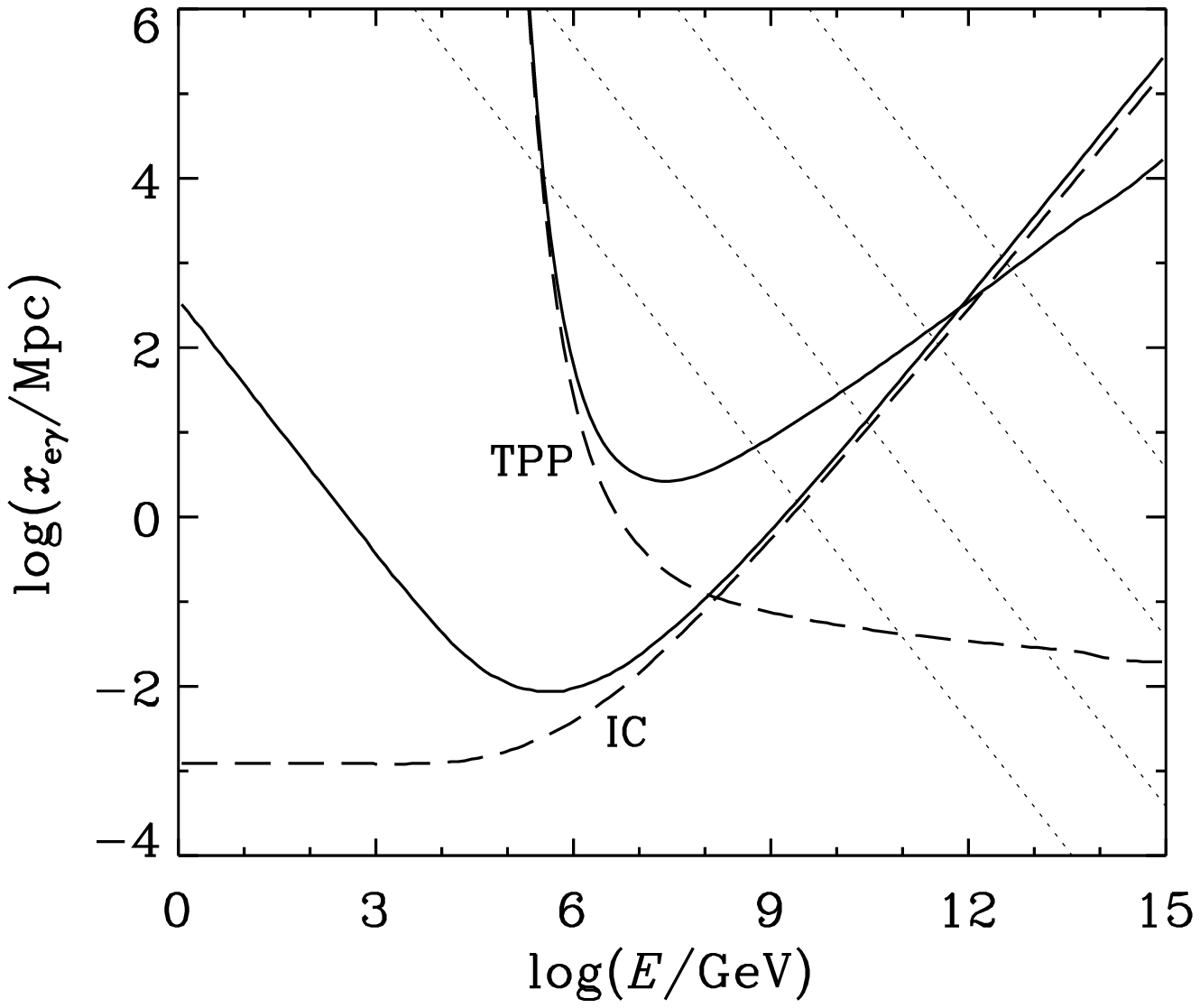,height=8cm}
\caption{(a) The mean interaction length for pair production for
$\gamma$-rays in: radio background calculated by Protheroe and
Biermann 1996 (solid curves labelled R), radio background of
Clark (1970) (dotted line); CMBR (2.7K), infrared and optical
background (IR) (Malkan and Stecker 1998).  The mean interaction
length for double pair production (4e) in the CMBR (Protheroe and
Johnson 1995) are also shown.  (b) The mean interaction length
(dashed line) and energy-loss distance (solid line), $E/(dE/dx)$,
calculated by Protheroe and Johnson (1995) for electron-photon
triplet pair production (TPP) and inverse-Compton scattering (IC)
in the CMBR.  The energy-loss distance for synchrotron radiation
is also shown (dotted lines) for intergalactic magnetic fields of
$10^{-9}$ (bottom), $10^{-10}$, $10^{-11}$, and $10^{-12}$ gauss
(top).
\label{fig:ggee_radio_e3kmfp}}
\end{figure}

There are several cascade processes which are important for UHE
CR propagating over large distances through a radiation field:
protons interact with photons resulting in pion production and
pair production; electrons interact via inverse-Compton
scattering and triplet pair production, and emit synchrotron
radiation in the intergalactic magnetic field; $\gamma$-rays
interact by photon-photon pair production.  Energy losses due to
cosmological redshifting of high energy particles and
$\gamma$-rays can also be important, and the cosmological
redshifting of the background radiation fields means that energy
thresholds and interaction lengths for the above processes also
change with epoch (see e.g. Protheroe et al.\ 1995).

The energy density of the extragalactic background radiation is
dominated by the CMBR.  Other components of the extragalactic
background radiation are discussed in the review of Ressel and
Turner (1990).  The extragalactic radiation fields which are
important for cascades initiated by UHE cosmic rays include the
cosmic microwave background, the radio background and the
infrared--optical background.  The radio background was measured
over thirty years ago (Bridle 1967, Clarke et al. 1970), but the
fraction of this radio background which is truly extragalactic,
and not contamination from our own Galaxy, is still debatable.
Berezinsky (1969) was first to calculate the mean free path on
the radio background.  More recently Protheroe and Biermann
(1996) have made a new calculation of the extragalactic radio
background radiation down to kHz frequencies.  The main
contribution to the background is from normal galaxies and is
uncertain due to uncertainties in their evolution.  The mean free
path of photons in this radiation field as well as in the
microwave and infrared backgrounds is shown in
Fig.~\ref{fig:ggee_radio_e3kmfp}(a).

Inverse Compton interactions of high energy electrons and triplet
pair production can be modelled by the Monte Carlo technique
(e.g. Protheroe 1986, Protheroe 1990, Protheroe et al.\ 1992,
Mastichiadis et al.\ 1994), and the mean interaction lengths and
energy-loss distances for these processes are given in
Fig. \ref{fig:ggee_radio_e3kmfp}(b).  Synchrotron losses must
also be included in calculations and the energy-loss distance has
been added to Fig. \ref{fig:ggee_radio_e3kmfp}(b) for various
magnetic fields.

Where possible, to take account of the exact energy dependences
of cross-sections, one can use the Monte Carlo method.  However,
direct application of Monte Carlo techniques to cascades
dominated by the physical processes described above over
cosmological distances takes excessive computing time.  Another
approach based on the matrix multiplication method has been
described by Protheroe (1986) and developed in later papers
(Protheroe and Stanev 1993, Protheroe and Johnson 1995).  A Monte
Carlo program is used to calculate the yields of secondary
particles due to interactions with radiation, and spectra of
produced pions are decayed to give yields of $\gamma$-rays,
electrons and neutrinos.  For the pion photoproduction
interactions a new program called SOPHIA is available (M\"{u}cke
et al. 1998).

%%%%%%%%%%%%%%%%%%%%%%%%%%%%%%%%%%%%%%%%%%%%%%%%%%%%%%%%%%%%%%%%%%%%%%%%%%%%%%%
\section{Radio Galaxies and Active Galactic Nuclei} 

\begin{figure}[htb]
\epsfig{file=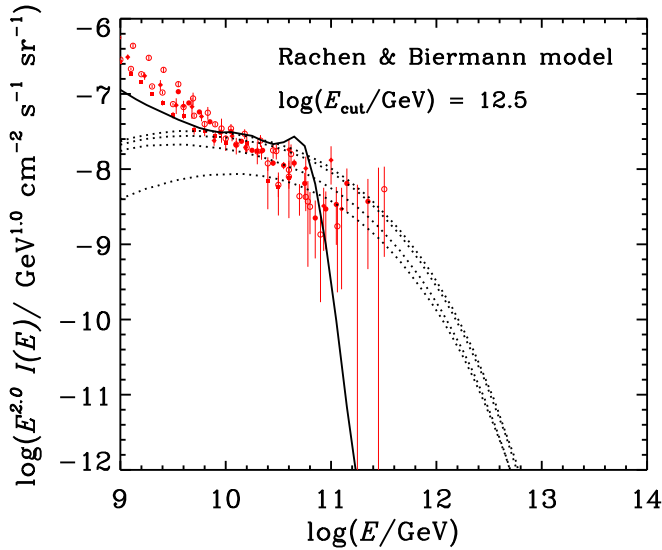,height=8cm}
\caption{Cosmic ray proton intensity multiplied by $E^{2}$ in the
model of Rachen and Biermann (1993) as calculated by Protheroe
and Johnson (1995) for proton injection up to $3 \times 10^{12}$
GeV (solid line).  Also shown are intensities of neutrinos
(dotted lines, $\nu_\mu , \bar{\nu}_\mu, \nu_e, \bar{\nu}_e$ from
top to bottom), and photons (long dashed lines).  Data are from
Gaisser and Stanev (1998).
\label{fig:rachen_model}}
\end{figure}

Rachen and Biermann (1993) have demonstrated that cosmic ray
acceleration hotspots of giant radio lobes of Fanaroff-Riley
Class II radio galaxies can fit the observed spectral shape and
the normalization at 10 -- 100 EeV to within a factor of less
than 10.  Protheroe and Johnson (1995) repeated Rachen and
Biermann's calculation to calculate the flux of diffuse neutrinos
and $\gamma$-rays which would accompany the UHE cosmic rays as a
result of pion photoproduction on the CMBR, and their calculated
flux is shown in Fig.~\ref{fig:rachen_model}.  The flux of
extremely high energy neutrinos may give important clues to the
origin of the UHE cosmic rays (for reviews of high energy
neutrino astrophysics see Protheroe 1998 and Learned and Mannheim
2000).  They may even be able to produce the observed UHE CR
above the GZK threshold through interacting with cosmological
neutrinos in our galactic halo as discussed in the next section
on ``Z-bursts''.

AGN jets may also accelerate protons to ultra high energies and
produce neutrino, gamma-ray and cosmic ray signals as a result of
pion photoproduction interactions in the intense AGN radiation
fields.  There are different versions of these models in which
the target photons are produced inside the blob, e.g.\ as
synchrotron emission by a co-accelerated population of electrons
(Mannheim 1993, 1995), or are external to the jet, e.g.\ from an
accretion disk (Protheroe 1997).  In addition, proton synchrotron
blazar models in which the high energy part of the spectral
energy distribution is mainly due to synchrotron radiation by
protons have been proposed for some blazars (M\"{u}cke and
Protheroe 2000, M\"{u}cke et al.\ 2001, Aharonian 2000).  The
relative contributions of various classes of AGN to the neutrino,
gamma-ray and cosmic ray backgrounds are discussed by Mannheim et
al.\ (2001).  

Interestingly, Tinyakov \& Tkachev (2001b, 2003) have claimed a
correlation between the arrival directions of cosmic rays, having
energies above 240~EeV from the Yakutsk array and above 480~EeV
from the AGASA array, with the the directions of the 22 most
powerful BL~Lac objects with redshifts $z$$>$0.1.  In this
analysis, the energy cuts were those for which there was an
indication of small-angle clustering from their previous analysis
(Tinyakov \& Tkachev 2001a).  As pointed out by Evans, Ferrer \&
Sarkar (2003) the redshift cut implies that cosmic rays from
these sources would be strongly affected by the GZK cut-off.
Furthermore, cosmic rays propagating to Earth from large
redshifts through the extragalactic magnetic field and CMBR may
have difficulty in reaching us in a Hubble time as well as being
severely affected by photoproduction interactions.  Evans, Ferrer
\& Sarkar (2003) fail to find a statistically significant
correlation between the cosmic ray arrival directions and the
directions of BL~Lac objects, and claim the correlation found by
Tinyakov \& Tkachev (2001b, 2003) to be spurious, and due to the
cuts imposed on the data.  We would add that there is no reason
to limit searches to BL~Lac objects as these objects are thought
to be Fanaroff Riley Class I radio galaxies having jets closely
aligned to our line of sight -- unless the particles being
searched for are neutral particles produced and relativistically
beamed in the jets, there is no reason to favour BL~Lac objects
over the more numerous Fanaroff Riley Class I radio galaxies.
The redshift cut ($z$$>$0.1) was used to ensure the BL~Lac
objects were powerful, but more powerful Fanaroff Riley Class I
radio galaxies can appear less luminous than less powerful BL~Lac
objects, due to the absence of strong relativistic beaming of
photons toward the observer.  Excluding nearer BL~Lac objects and
Fanaroff Riley Class I radio galaxies seems to us to be
illogical.  Indeed, the sub-parsec scale jets of the nearby
Fanaroff-Riley Class I radio galaxy M87, near the centre of the
Virgo Cluster, may be a mis-aligned blazar of the BL~Lac type,
and may accelerate protons to ultra-high energies which could
propagate to Earth if the magnetic field topology between our
Galaxy and the Virgo Cluster were favourable (Protheroe, Donea
and Reimer 2003).

Quasars are also an interesting possibility.  Farrar and Biermann
(1998) found a correlation between the arrival directions of the
five highest energy cosmic rays having well measured arrival
directions and radio-loud flat spectrum radio quasars with
redshifts ranging from 0.3 to 2.2.  The probability of obtaining
the observed correlation by chance is estimated to be 0.5\%.
Although the statistical significance is not overwhelming, and
indeed other researchers find no statistically significant
evidence for such a correlation (Hoffman 1999, Sigl et al.\
2001), if further evidence is provided for this correlation the
consequences would be far reaching.  The distances to these AGN
are far in excess of the energy-loss distance for pion
photoproduction by protons.  Furthermore, given the existence of
intergalactic magnetic fields, any charged particle would be
significantly deflected and there should be no arrival direction
correlation with objects at such distances.  Hence, the particles
responsible would need to be be stable, neutral, and have a very
low cross section for interaction with radiation.  Of currently
known particles, only neutrinos fit this description, however
supersymmetric particles are another possibility, as suggested by
Farrar and Biermann (1998), although this possibility now appears
to be ruled out (Gorbunov, Raffelt \& Semikoz 2001).

\subsection{Z-Bursts}

The difficulty of having high energy neutrinos producing the
highest energy cosmic rays directly is circumvented if the
neutrinos interact well before reaching Earth and produce a
particle or particles which will produce a normal looking air
shower.  As suggested by Weiler (1999), this may occur due to
interactions with the 1.9~K cosmic background neutrinos (see also
Gelmini and Kusenko 2000).  The clustering of relic neutrinos in
hot dark matter galactic halos would give an even denser nearby
target for ultra high energy neutrinos as suggested by Fargion et
al. (1999).  The cross section is much larger for resonant $Z^0$
production which would occur for a UHE neutrino with energy $\sim
m_Z^2/2m_\nu \sim 4 \times 10^{21}/(m_\nu/$eV)~eV.  From the
recent SNO results( Ahmad et al., 2002), the Super-Kamiokande
atmospheric neutrino results (Toshito et al., 2001) and the
tritium $\beta$ decay results (Bonn et al., 2001), Ahmad et al.\
(2002) concluded that the sum of mass eigenvalues of oscillating
neutrinos was in the range 0.05--8.4~eV.  Hence, the required UHE
neutrino energy for resonant $Z^0$ production is very
interestingly at or above the energies of UHE CR near the GZK
cut-off.  In this ``Z-Burst'' scenario, the $Z^0$ produced would
have a comparable energy to the UHE neutrino, and decay into
leptons and hadrons (including nucleons) which could be detected
as UHE CR.  In principle, there is also the possibility of the
determination of absolute neutrino masses from Z-bursts if the
galactic cosmic ray spectrum from normal acceleration were known
(P\"as and Weiler 2001, Fodor et al.\ 2002).

The main problem with the Z-Burst scenarios is that except for
the case of unrealistic source models (production of UHE
neutrinos with very few other UHE particles) or rather extreme
over-densities (by $>$$10^3$) of relic neutrinos, the cascade
gamma-ray flux would exceed the GeV gamma-ray intensity observed
by EGRET (e.g.\ Kalashev et al.\ 2002).  Also, extremely high
fluxes of UHE neutrinos are required to explain the observed UHE
CR spectrum.  They are well in excess of those expected from AGN,
and the possibility of explanation in terms of $X$ particle decay
exclusively into neutrinos seems unsatisfactory (Berezinsky et
al.\ 2002).  Nevertheless, these fluxes are in principle
detectable with existing neutrino detectors, and if they exist
should certainly be detected with future large area cosmic ray
detectors, such as the Pierre Auger Observatory, which are also
sensitive to UHE neutrinos.  If detected, the observed UHE
neutrino flux, together with the observed UHE CR flux and
anisotropy would place important constraints on the mass of the
possible neutrino species forming the dark matter galactic halo,
and its radial distribution (e.g.\ Singh and Ma 2003) as well as
the source of the UHE neutrinos.  Their detection would also
require a re-evaluation of the our understanding of
electromagnetic and hadronic cascading in the CMBR and other
radiation fields.

\section{Topological Defects}

Topological defects (TD) such as monopoles, cosmic strings,
monopoles connected by strings, etc., may be produced at the
post-inflation stage of the early Universe.  In the process of
their evolution the constituent superheavy fields (particles) may
be emitted through cusps of superconducting strings, during
annihilation of monopole-antimonopole pairs, etc.  These
particles, collectively called X-particles, can be superheavy
Higgs particles, gauge bosons and massive supersymmetric (SUSY)
particles.  These are generally very short-lived, and their decay
followed by a hadronization cascade could produce an observable
signal.  Signals of TD origin would be affected by
interactions/cascading during propagation over cosmological
distances to Earth.  Protheroe and Johnson (1996) pointed out the
importance of including pair-synchrotron cascades in UHE CR
propagation and, following their approach, Protheroe and Stanev
(1996) showed that the $\gamma$-ray flux for many TD models of
UHE CR exceeded that observed at 100~MeV energies for $B \ge
10^{-9}$ G.

There could be also superheavy quasi-stable particles with
lifetimes larger (or much larger) than the age of the Universe.
These particles could be produced by many mechanisms during the
post-inflation epoch, and survive until the present epoch.  One
interesting process is the ``gravitational production of
super-heavy particles'', in which no interaction of X-particles
is required.  Also, string theories predict the existence of
other super-heavy particles (``cryptons'') which are metastable
and could in principle form part of the cold dark matter (CDM)
(see, e.g., Kolb 1998, Ellis 2000).  As with any other kind of
CDM, super-heavy quasi-stable X-particles would cluster in
galactic halos. The same clustering would also occur for some TD,
such as monopolonium, monopole-antimonopole pairs connected by a
string, and vortons.  Cosmic ray signals from all these objects
would reach us relatively attenuated.  Perhaps the most promising
WIMP CDM candidate is the lightest SUSY particle (LSP) with mass
only 20--1000 GeV, and so would not produce UHE CR.

\subsection{Fragmentation functions}

TD such as cosmic strings, necklaces, etc., are extragalactic,
and could produce an extragalactic signal through the decay of
short-lived X-particles.  TD which accumulate in galaxy halos
(monopolonia, monopole-antimonopole-pairs and vortons) could
produce a galactic signal through the annihilation/emission and
decay of short lived X-particles which would in turn decay
promptly into Standard Model (SM) states.  Super-heavy
quasi-stable particles ($\tau \gg t_0$) would decay similarly,
and also be clustered as CDM in galactic halos.

The $X$ particle decay products annihilate and could give
rise to a jet of hadrons, e.g.
\[
 X  \to
\left\{ \begin{array}{c} W^+W^- \\ Z^0Z^0 \\ \bar{q}q \\ e^+e^- \\
{\rm etc.} \end{array} \right\} \to  \mbox{~2 jets} \to \left\{ 
\begin{array}{l}\gamma{\rm -rays}  \\ 
{\rm neutrinos}   \\ \mbox{nucleons ($\sim 5\%$)}  \\ {\rm electrons} 
\end{array} \right.
\]
Energy spectra of the emerging particles, the ``fragmentation
functions'', were first calculated by Hill (1983).  Each jet has
energy $m_Xc^2/2$, and so one defines a dimensionless energy for
the cascade particles, $x=2E/m_Xc^2$.  The fragmentation function
for ``species a'' is then defined as $dN_a/dx$.  A very flat
spectrum of particles results, and extends up to $\sim m_Xc^2/2$.
In the case of decay of CDM in galactic halos, the resulting UHE
CR spectrum is proportional the fragmentation function for
nucleons.

\begin{figure}[htb!] % 
\centerline{\epsfig{file= 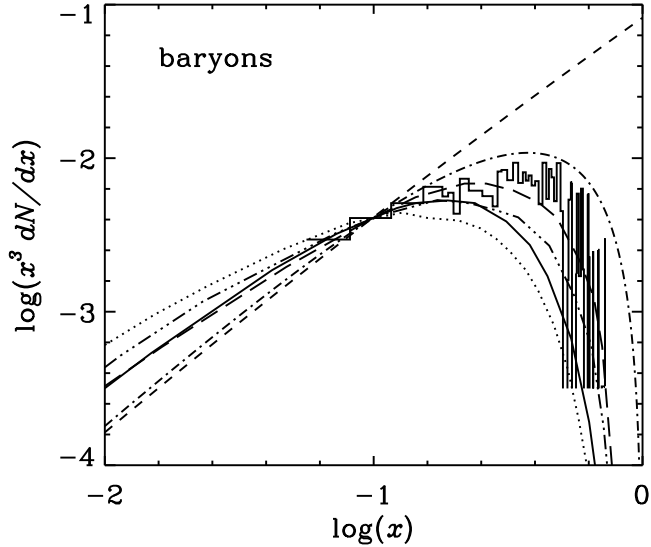,width=4.5in}}
\vspace{10pt}
\caption{Fragmentation functions for hadronization of baryons
normalized at $x=0.1$ to the recent QCD calculation of Sarkar and
Toldra (2002) for $m_X=10^{12}$GeV (solid curve): Hill(1983)
(dot-dashed curve); Berezinsky et al.\ (1997) MLLA approximation
(short dashed curve); Monte Carlo results of Birkel and Sarkar
(1998) for $m_X=10^{11}$GeV (solid histogram); Berezinsky and
Kachelriess (2000) for $m_X=10^{12}$GeV SUSY-QCD (long dashed
curve); Sarkar and Toldra (2002) for $m_X=10^{12}$GeV SUSY-QCD
(dotted curve); Rubin (2000) $m_X=10^{12}$GeV QCD
(dot-dot-dot-dashed curve). }
\label{fig1}
\end{figure}

Some recent calculations of the fragmentation functions used the
Modified Leading Logarithm Approximation (MLLA) which is valid
only for $x \ll 1$, and in more recent QCD calculations
PYTHIA/JETSET (Singh and Ma 2003) or HERWIG Monte Carlo event
generators were used.  The fragmentation functions due to Hill
(1983), and those of Berezinsky et al.\ (1997) based on the MLLA
are compared in Fig.~\ref{fig1}.  Initially, the inclusion of the
production of SUSY particles was done by putting 40\% of the
cascade energy above threshold for production of SUSY particles
into LSP (Berezinsky and Kachelriess 1998), thereby steepening
the fragmentation functions for normal particles at high energy.
Birkel and Sarkar (1998) showed that even without inclusion of
SUSY production there is a significant dependence on $m_X$, such
that for high $m_X$ the fragmentation functions are steeper, as a
direct consequence of the well-known Feynman scaling violation in
QCD.  Fragmentation functions calculated by Birkel and Sarkar
(1998) using the HERWIG event generator have been added to
Fig.~\ref{fig1}, but this event generator is now known to
overestimate production of nucleons by a factor
$\sim$2--3,(Sarkar~2000, Rubin~2000).  Recent calculations by
Rubin~(2000), Sarkar and Toldra (2002) and Berezinsky and
Kachelriess (2000) (added to Fig.~\ref{fig1}) have used improved
treatments of SUSY particle production, and result in only
$\sim5$--12\% of the cascade energy going into LSP.  Very
recently, Barbot \& Drees (2003) have produced a complete set of
fragmentation functions for any SUSY particle of the minimal
supersymmetric extension of the standard model into protons,
photons, electrons, neutrinos and the LSP.

\begin{figure}[htb!] % fig 3
\hspace*{-5em}\epsfig{file=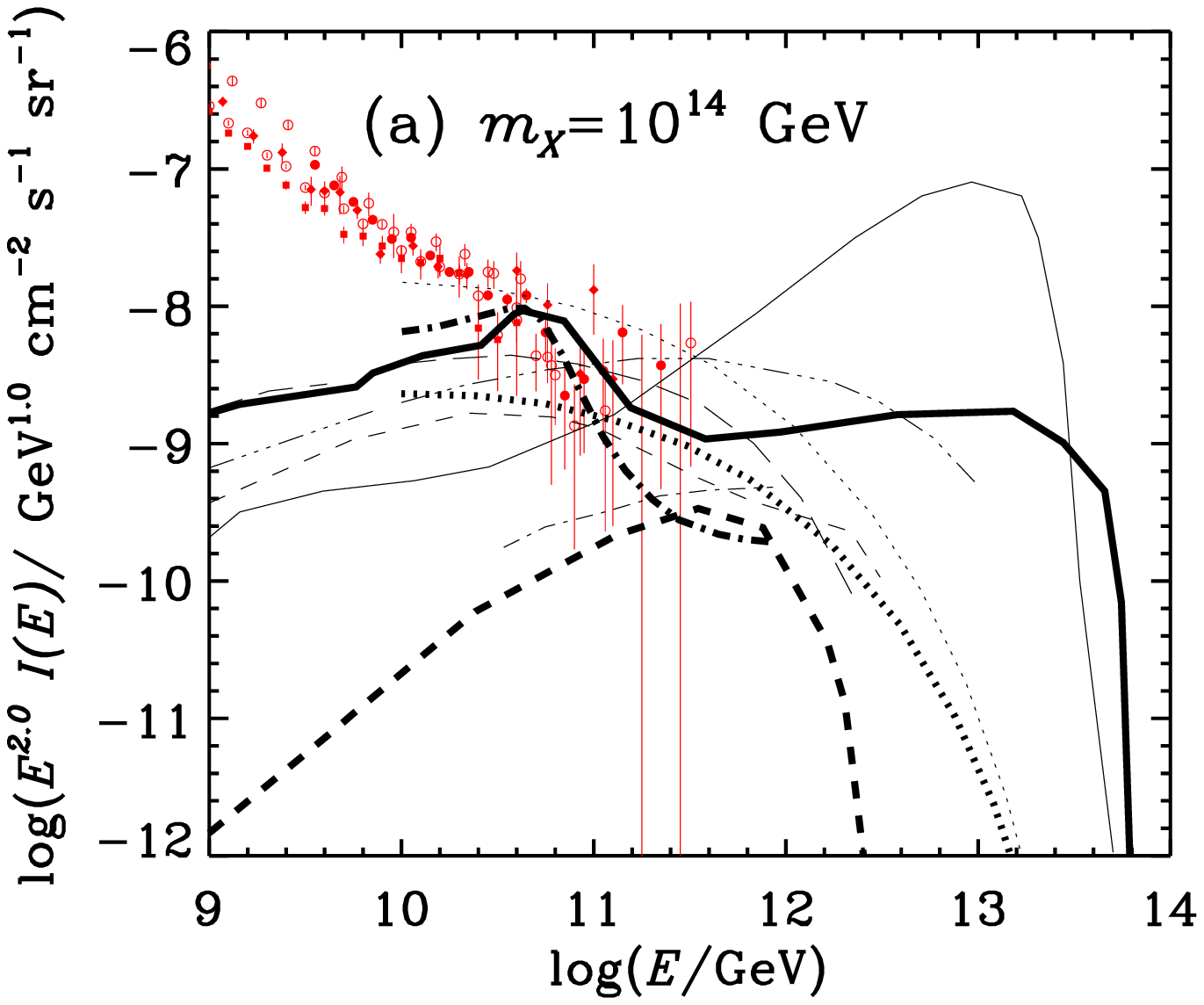,height=8cm}\hspace*{-6em}\epsfig{file=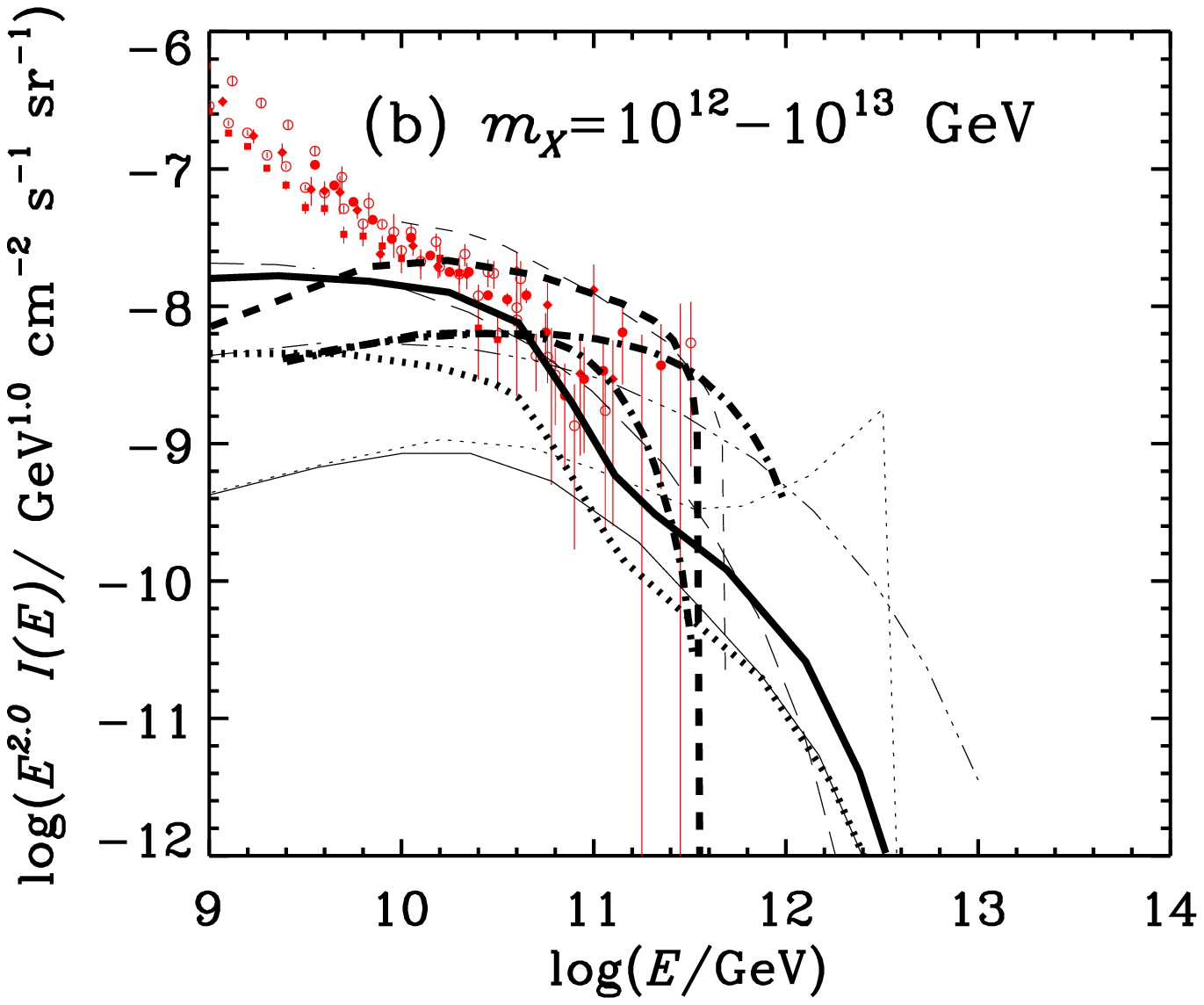,height=8cm}
\vspace{10pt}
\caption{Predicted TD model spectra of protons (thick curves) and
cascade photons (thin curves) compared with cosmic ray data from
Gaisser and Stanev (1998).  (a) $m_X = 10^{14}$ GeV: Protheroe \&
Stanev (1996) (solid curves); Sigl et al.\ (1999) $X \to \nu\nu$
(short dashed curves); Blasi (1999) super-heavy relic halo
population, SUSY-QCD and $\pi\mu e$ synchrotron (gamma rays only,
long dashed curve), QCD and $\pi\mu e$ synchrotron (gamma rays
only, dot-dot-dot-dashed curve); Berezinsky et al. (1998)
super-heavy relic halo population, SUSY-QCD (dotted curves),
necklaces SUSY-QCD (dot-dashed curves).  (b) $m_X =
10^{12}$--$10^{13}$ GeV: Sigl et al. (1999) $m_X=10^{13}$~GeV, $X
\to q+q$ QCD (solid curves), $m_X=10^{13}$~GeV, $X \to q+l$ QCD
(dotted curves); Blasi (1999) $m_X=10^{13}$~GeV super-heavy relic
halo population, SUSY-QCD and $\pi\mu e$ synchrotron (gamma rays
only, long dashed curve), QCD and $\pi\mu e$ synchrotron (gamma
rays only, dot-dot-dot-dashed curve); Sarkar and Toldra (2002)
super-heavy relic halo population QCD best fit $m_X=10^{12}$~GeV
(protons only, leftmost dot-dashed curve) and SUSY-QCD best fit
$m_X=5 \times 10^{12}$~GeV (rightmost dot-dashed curve).  }
\label{figTD}
\end{figure}

\subsection{Viability of Dark Matter Origin of UHECR}

Predictions for some TD and massive relic particles are show in
Fig.~\ref{figTD}.  If CDM consists of particles associated with
TD distributed uniformly throughout the Universe, then UHE CR are
subject to the GZK cut-off.  In this case $\gamma$-ray signals
result from a pair-synchrotron cascade in background radiation
and extragalactic magnetic fields.  The magnetic fields used in
some cascade calculations may have been unrealistically low, and
it does appear difficult to explain the super-GZK events with
such TD models without the flux of cascade gamma-rays exceeding
the observed 100~MeV gamma-ray background.

Most of the matter in the Universe is CDM, and if it consists of
massive relic particles they should cluster in galaxy halos. In
this case, decay of massive relic particles would produce UHECR
signals weakly anisotropic toward the GC, and the UHE CR spectrum
would not have a GZK cut-off.  Of the ``top down'' scenarios,
these models currently seem to us to be the most viable.  See
Sarkar and Toldra (2002) for recent work on decay of superheavy
dark matter particles.

%%%%%%%%%%%%%%%%%%%%%%%%%%%%%%%%%%%%%%%%%%%%%%%%%%%%%%%%%%%%%%%%%%%%%%%%%%%

\section{Propagation through Magnetic Fields}

Cosmic rays reach us after travelling through the magnetic fields which 
pervade space.  Details of the strength and structure of such fields are 
unknown but broad generalizations are possible within certain volumes
of the Universe.

Our galaxy is of spiral structure and the galactic magnetic field
has a regular component with a characteristic strength of the
order of microgauss and which seems to be associated with the
spiral arms.  Additionally, there is a turbulent, random,
component which is at least as strong as the regular component.
This component is even less well known since measurements of
Faraday rotation or other techniques tend to average out when the
line of sight transits a number of turbulence cells.  The
spectrum of turbulence scales is often assumed to be of a
Kolmogorov kind which has the important property of being
dominated by the largest scale sizes.  Within our Galaxy, the
largest internal structures tend to be of 100~pc scales
(e.g. supernova remnants).  With a field strength of a few
microgauss, this means that significant scattering of cosmic rays
will occur at least to a few times $10^{17}$eV since a proton
with this energy has a gyroradius of 100~pc in a 1~$\mu$G field.
Hence, the largest scale lengths in the turbulence of the
Galactic magnetic field tend to dominate the propagation of the
(more energetic) UHE CR in the Galaxy.

The particles follow paths rather like random walks up to the
scale size of the turbulence.  A first approximation to galactic
propagation is then diffusion.  Honda (1987) has given an
excellent discussion of extensions to make the simple picture
more realistic.  That work, and similar propagation modelling by
Clay (2000), gives us some understanding of resulting measurable
properties of the cosmic ray beam.  Since the propagation is
diffusive, the time for a particle to leave the galaxy is greater
than the simple direct transit time.  This containment time
increase results in an increase in flux over that which would
have been observed had there only been straight line propagation
from galactic sources.  Containment time calculations thus allow
us to crudely determine a 'source spectrum'.  In one recent
analysis (Clay 2000), the source spectrum shows no knee and,
possibly, no ankle.  It may be that both those features (and
certainly the knee) are consistent with purely propagation
effects.  The resulting source spectrum is a power law with an
index of 2.  There is a problem with such an explanation for the
ankle since particles above that energy travel in rather straight
lines and a 'Milky Way' perhaps ought to be visible in the
anisotropy data.  However, data at these energies are somewhat
sparse and, of course, the AGASA/SUGAR source (Hayashida et al.\
1999, Bellido et al.\ 2001) could be just such an effect.

\begin{figure}[htb]
\epsfig{file=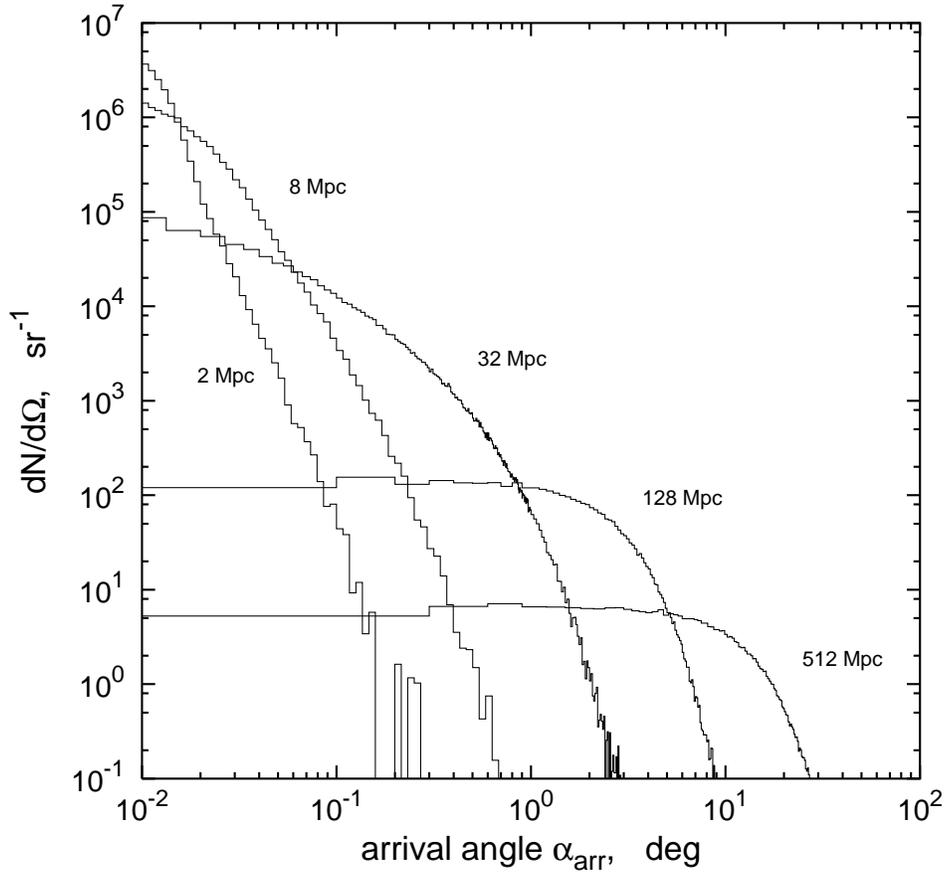,height=12cm}
\caption{Angular distribution of the arrival angle at Earth for
mono-energetic injection of protons of energy $E=10^{21.5}$~eV,
and for various source distances as indicated (from Stanev et
al.\ 2000).}
\label{fig:Stanevetal2000_angdist}
\end{figure}

In considering the propagation from extragalactic sources, there
are two environments to consider.  They are the intra-cluster
magnetic fields of both the source galaxy and our own galaxy, and
the inter-cluster field.  Clarke et al.\ (2001) have shown that,
remarkably, a characteristic intra-cluster magnetic field
strength in a rich galactic cluster fills the cluster and has
microgauss strengths -- maybe $5\mu$G in the inner 500~kpc.  If
we make a first approximation to a diffusion coefficient to be a
factor $\eta \! > \! 1$ times the minimum (Bohm) diffusion
coefficient, i.e.\ $\eta r_gc/3$, we can assume diffusive
propagation and derive an estimate of the time to reach a given
root-mean-square displacement.  If we consider a $10^{19}$eV
particle in the $5\mu$G field, we find a required time of
$10^8/\eta$~yr just to leave the source cluster through 500~kpc.
The GZK effect is clearly relevant here.  If the source is in the
Virgo Cluster of galaxies, a cosmic ray must then travel through
intercluster space and then reach us though our own cluster
field.  It may be that the intercluster field is also at
significant levels.  In this case, we are looking at tens of
megaparsec from the nearest likely AGN source with a transit
time, being dependent on the square of the distance, and becoming
greater than the age of the Universe.  This is clearly an issue
which pushes us to a careful consideration of very local sources.

This argument is rather crude.  Sigl (2000) has modelled time
delays for particles travelling 10~Mpc in a turbulent $0.3\mu$G
field.  Even for that modest field strength, transit times of
$10^8$~yr apply at 10--100~EeV.  Apart from any concern about
particles reaching us within the age of the Universe, our
comments on diffusion times emphasize that it may not make sense
to correlate cosmic ray observations with sources beyond 10~Mpc
unless one can be sure that those sources have a lifetime for
emission substantially greater than $10^8$yr.  Monte Carlo
calculations have been made by Stanev et al.\ (2000) for
propagation through an irregular magnetic field having a
Kolmogorov spectrum of turbulence with minimum wavenumber
1~Mpc$^{-1}$ and energy density equal to that of a much smaller
1~nG intergalactic field.  Propagation from sources at several
distances take account of diffusion in this turbulent field as
well as interactions with the CMBR using the SOPHIA event
generator (M\"{u}cke et al.\ 2000) and redshifting, and give the
distributions in energy, time delay relative to straight-line
propagation, and angular distribution about source direction.
Fig.~\ref{fig:Stanevetal2000_angdist} shows the angular
distribution of protons with initial energies 300~EeV arriving
from sources at distances 2~Mpc, 8~Mpc, \dots 512~Mpc away.

\begin{figure} 
\centerline{\epsfig{file=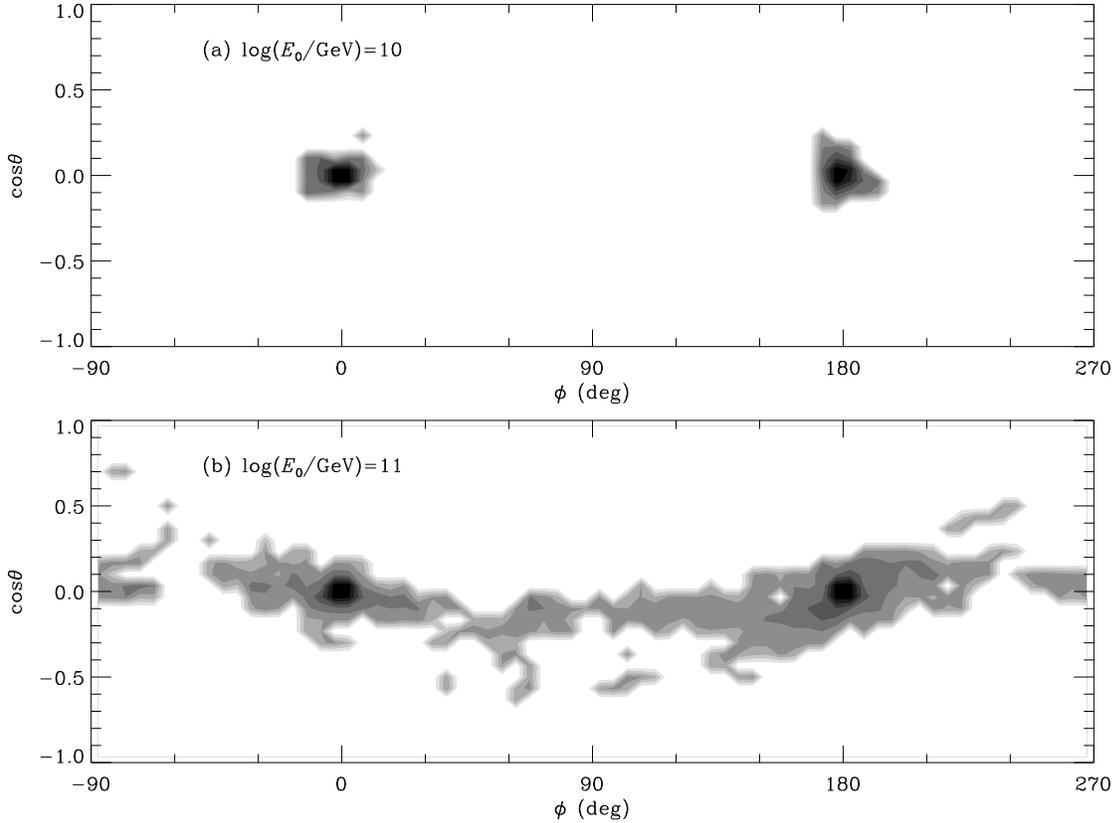,width=\hsize}} 
\caption{Enhancement factor relative to rectilinear propagation
as a function position $(\cos\theta,\phi)$ a sphere of radius $\sim
16$~Mpc centred on the origin for protons of initial energy (a)
$E_0=10^{19}$~eV and (b) $E_0=10^{20}$~eV injected isotropically
at the origin in the wall/void model of the IGMF discussed in the
text.  Positions $(0,0^\circ)$ and $(0,180^\circ)$
are on the field line threading M87 (from Protheroe et al.\
2003).}
\label{fig:m87cr}
\end{figure}

The situation is quite different if the intergalactic magnetic
field structure is based on the observation of microgauss fields
in clusters of galaxies (Kronberg 1994), and of clusters
occurring in networks of ``walls'' separated by ``voids''.  Using
a wall/void model similar to that of Medina Tanco (1998),
Protheroe et al.\ (2003) discuss propagation from M87 (located
close to the centre of the Virgo Cluster) assuming that it and
our galaxy are embedded in the same wall (thickness 2.5~Mpc)
which has a regular magnetic field of $10^{-7}$~G in the plane of
the wall and $10^{-10}$~G in the surrounding void, and an
irregular component with 30\% of the energy density of the
regular component and having a Kolmogorov spectrum of turbulence.
Modelling M87 as a mis-aligned BL~Lac object, they found the
UHECR output from M87 to be at a level such that if UHECRs
travelled in straight lines they would give an average intensity
at Earth a factor $\sim$20 below that observed.  However,
observed fluxes will be very different for propagation in a more
realistic magnetic field structure such as that described above.
Propagation results for the wall/void model are shown for two
initial energies $10^{19}$~eV and $10^{20}$~eV in
Fig.~\ref{fig:m87cr}.  Note that Fig.~\ref{fig:m87cr} does {\em
not} give the anisotropy that would be observed at the Earth due
to cosmic rays from M87, but rather the enhancement factor
(relative to straight line propagation) of the cosmic ray flux
from M87 at positions on a sphere of radius 16~Mpc centred on
M87.  Enhancement factors $\sim$$10^3$ exist if the Earth is
within $\sim$$1.5$~Mpc of a field line originating at M87. If
this (rather special) condition were met M87 could easily explain
the observed UHECR.  Because of its very high black hole mass M87
was probably much more active at earlier times than at present
(many objects exhibit a high state for $\sim$5\% of the time)
which makes M87 a more attractive candidate source of the UHECR.

%%%%%%%%%%%%%%%%%%%%%%%%%%%%%%%%%%%%%%%%%%%%%%%%%%%%%%%%%%%%%%%%%%%%%%%%%%%%%

\section{Conclusions}

Cosmic ray astrophysics at the highest energies is entering a new
era with major new facilities being proposed and developed.  The
tiny flux of particles in this range of energies challenges our
understanding of the astrophysics which we apply to many other
studies.  We are led to examine processes in extreme regions such
as the various components of AGN, to ask how particle
acceleration may take place there, and to ask for detailed
information on photon and magnetic fields throughout our local
universe.  This challenging work now truly complements studies
throughout photon astrophysics.

%%%%%%%%%%%%%%%%%%%%%%%%%%%%%%%%%%%%%%%%%%%%%%%%%%%%%%%%%%%%%%%%%%%%%%%%%%%%%

\section*{Acknowledgments}

Our research is supported by grants from the Australian Research
Council.  Our understanding of issues related to TD/CDM origin
models has benefitted from discussions with Venya Berezinsky and
Subir Sarkar.  We thank the referees for helpful comments.

\section*{References}

\reference Abu-Zayyad, T.\ et al. 2001, ApJ, 557, 686

\reference Abu-Zayyad, T.\ et al. 2003, Submitted to Astropart.\ Phys., astro-ph/0208301

\reference Achterberg, A., Gallant, Y.A., Kirk, J.G., Guthmann, A.W. 2001, MNRAS, 328, 393

\reference Aharonian, F.A., Bhattacharjee P., \& Schramm D.N. 1992, Phys.\ Rev.\ D, 46, 4188

\reference Aharonian, F.\ 2000, New Astron., 5, 377

\reference Ahmad, Q.R., et al., 2002, Phys.\ Rev.\ Lett., 89, 011301; Phys.\ Rev.\ Lett., 89, 011302 

\reference Anchordoqui, L.A.,  Dova, M.T., Epele, L.N., \&  Swain, J.D.  1998, Phys.\ Rev.\ D,  57, 7103

\reference Auger Collaboration Contribution 2001, in Proceedings of XXVII Cosmic Ray Conference, Hamburg, Edited by M. Simon et al., Copernicus Gesellschaft 2001, Vol. 2, pp 699-787

\reference Axford, W.I., Lear, E., \& Skadron, G. 1977, in Proc.\ 15th Int.\ Cosmic Ray Conf., Plovdiv vol.\ 11, p 132

\reference Barbot, C.\& Drees, M., 2003, Astropart.\ Phys., 20, 5

\reference Baltrusaitis, R.M.\ et al. 1985, Nucl.~Instr.~Meth., A240, 410

\reference Baring, M.G. 1997, in Proc.\ of XXXIInd Rencontres de Moriond, ``Very High Energy Phenomena in the Universe'', eds.\ Giraud-Heraud, Y. \& Tran Thanh Van, J., (Editions Frontieres, Paris, 1997), p. 97

\reference Bednarek, W., \& Protheroe, R.J., 2002, Astropart.\ Phys., 16, 397

\reference Bednarz, J., \& Ostrowski, M. 1996, MNRAS, 283, 447

\reference Bednarz, J., \& Ostrowski, M. 1998, Phys.\ Rev.\ Lett., 80, 3911

\reference Bednarz, J. 2000, MNRAS, 315, 37

\reference Bell, A.R. 1978, MNRAS, 182, 443

\reference Bell, A. R., \& Lucek, S. G. 1996, MNRAS, 283, 1083

\reference Bellido, J.A., Clay, R.W., Dawson, B.R., \& Johnston-Hollitt, M. 2001, Astropart.~Phys., 15, 167

\reference Berezhko, E.G., \& Krymski, G.F. 1988, Usp. Fiz. Nauk, 154, 49

\reference Berezhko, E. G. 1994, Astron.\ Lett., 20, 75

\reference Berezhko, E.G. 2001, Space Sci.\ Rev., 99, 295

\reference Berezhko, E.G. \& V\"olk, H.J. 2000, A\&A, 357, 283

\reference Berezinsky, V.S. 1970, Sov.\ J.\ Nucl.\ Phys.,  11, 222

\reference Berezinsky, V., Blasi, P., \& Vilenkin, A. 1998, Phys.\ Rev.\ D, 58, 103515

\reference Berezinsky, V., Kachelriess, M., \& Vilenkin A. 1997, Phys.\ Rev.\ Lett., 79, 4302

\reference Berezinsky, V., \& Kachelriess, M. 1998, Phys.\ Lett.\ B, 434, 61 

\reference Berezinsky, V., \& Kachelriess, M. 2001, Phys.\ Rev.\ D, 63, 034007

\reference Berezinsky, V., Kachelreiss, M., \& Ostapchenko, S. 2002, Phys.\ Rev.\ Lett., 89 171802

\reference Bhattacharjee, P. \&  Sigl, G. 2000, Phys.~Rep., 327, 109

\reference  Birkel, M., \& Sarkar, S. 1998, Astropart.\ Phys., 9, 297

\reference Blandford, R.D., \& Ostriker, J.P. 1978, ApJ, 221, L29

\reference Blandford, R, \& Eichler, D. 1987, Phys.\ Rep., 154, 1

\reference Blasi, P. 1999, Phys.\ Rev.\ D, 60, 123514 

\reference Blasi, P., Epstein, R. I., \& Olinto, A. V., 2000, ApJ, 533, L123

\reference Biermann, P.L., \&  Strittmatter, P.A. 1987, ApJ., 322,  643 

\reference Bird, D.J.\ et al. 1995, ApJ., 441, 144

\reference Bonn, J.\ et al. 2001, Nucl.\ Phys.\ B Proc.\ Suppl., 91, 273

\reference Bridle, A.H. 1967, MNRAS, 136, 14

\reference Casse, F., \& Markowith, A. 2003, A\&A, 404, 405

\reference Clarke, T.A., Brown, L.W. \& Alexander, J.K. 1970, Nature, 228, 847

\reference Clarke, T.E., Kronberg, P.P., \& Bohringer, H. 2001, ApJ, 547,  L111

\reference Clay, R.W., McDonough, M.-A.,  \& Smith, A.G.K. 1998, Pub.~Astron.~Soc.~Aust., 15, 208

\reference Clay, R.W. 2000, Pub.~Astron.~Soc.~Aust., 17, 212

\reference Clay, R.W. 2002,  Pub.~Astron.~Soc.~Aust., 19, 228

\reference de Gouveia Dal Pino, E. M., \& Lazarian, A. 2001, ApJ, 560, 358

\reference Dermer, C.D. 2002, ApJ, 574, 65

\reference Drury, L.O'C. 1983a, Space Sci.\ Rev., 36, 57

\reference Drury, L.O'C. 1983b, Rep.\ Prog.\ Phys., 46, 973

\reference Drury, L.O'C., Duffy, P., Eichler, D., \& Mastichiadis, A. 1999, A\&A, 347, 370

\reference Elbert, J.W., \&  Sommers, P. 1995, ApJ, 441, 151

\reference Ellis J. 2000, in Proc.\ of the 26th Int.\ Cosmic Ray Conf., AIP Conf.\ Proc.\ 516, Eds.\ B.L. Dingus et al. (American Inst. Physics), p 21

\reference Ellison, D.C, \& Eichler, D. 1984, ApJ, 286, 691

\reference Ellison, D.C, Jones, F.C., \& Reynolds, S.P. 1990, ApJ, 360, 702

\reference Ellison, D.C., Baring, M.G., \&  Jones, F.C. 1995, ApJ, 453, 873

\reference Ellison, D.C., \& Double, G.P. 2002, APh, 18, 213

\reference Epele, L.N., \&  Roulet, E. 1998, J.\ High Energy Phys., 9810, 9

\reference Evans, N.W., Ferrer, F., \& Sarkar, S., 2003, Phys.\ Rev., D67, 103005

\reference Falcke, H., \& Gorham, P. 2003, Astropart.\ Phys., 19, 477

\reference Fargion, D., Mele, B., \& Salis, A. 1999, ApJ, 517, 725

\reference Farrar, G.R.,  \& Biermann P.L. 1998, Phys.\ Rev.\ Lett., 81, 3579; erratum 1999, Phys.\ Rev.\ Lett., 83, 2478;  Phys.\ Rev.\ Lett., 83, 1999, 2472 (response to Hoffman 1999)

\reference Fodor, Z., Katz, S.D., \& Ringwald, A. 2002, JHEP, 0206 046

\reference Gaisser, T.K. 1990, ``Cosmic Rays and Particle Physics'', (Cambridge University Press, Cambridge, 1990).

\reference Gaisser T.K., \& Stanev T. 1998, Europhys.\ J., 3, 1

\reference Galama, T.J., et al. 1998, Nature, 395, 670

\reference Gallant, Y. A., Achterberg, A., \& Kirk, J. 1999, A\&AS, 138, 549

\reference Gelmini, G., \& Kusenko, A. 2000, PRL, 84 1378

\reference Gold, T., 1975 Phil.\ Trans.\ Roy.\ Soc., Series A, 277, 453

\reference Gorbunov, D.S., Raffelt, G.G.,  \& Semikoz, D.V., 2001, Phys.\ Rev., D64, 096005

\reference Greisen, K. 1966, Phys.\ Rev.\ Lett., 16, 748

\reference Halzen F., Vazquez R.A., Stanev T., \&  Vankov H.P. 1995, Astropart.\ Phys., 3, 151

\reference Haswell, C.A., Tajima, T., \& Sakai, J.-L. 1992, ApJ, 401, 495

\reference Hayashida, N.\ et al. 1999, Astropart.\ Phys., 10, 303

\reference Hill, C.T. 1983, Nucl.\ Phys.\ B, 224, 469

\reference Hillas, A.M. 1984, Ann.~Rev.~Astron.~Astrophys., 22, 425

\reference Hoffman, C.M. 1999,  Phys.\ Rev.\ Lett., 83, 2471 

\reference Honda, M. 1987, ApJ., 319, 836

\reference Jokipii, J.R., \& Morfill G.E. 1985, ApJ, 290, L1

\reference Jokipii, J.R. 1987, ApJ, 313, 842

\reference Jones, F.C., \& Ellison, D.C. 1991, Space Sci.\ Rev., 58, 259

\reference Jones, F.C. 1994, ApJS, 90, 561  

\reference Karakula, S., \&  Tkaczyk, W. 1993, Astropart.\ Phys., 1, 229

\reference Kawabata, K.S. et al.\ 2003, astro-ph/0306155

\reference Kirk, J.G., \& Schneider, P. 1987, ApJ,  315, 425

\reference Kirk, J.G., \& Duffy, P. 1999, JPhG, 25, 163

\reference Kirk, J.G., Guthmann, A.W., Gallant, Y.A., \& Achterberg, A. 2000, ApJ, 542, 235

\reference Kolb, E.W. 1998, in Proceedings of the NATO Advanced Study Institute on Techniques and Concepts of High-Energy Physics (10th: 1998: St. Croix, V.I.) Ed. T. Ferbel.

\reference Kronberg, P.P. 1994, Rep.\ Prog.\ Phys.\, 57, 325

\reference Krymsky, G.F. 1977, Dokl.\ Akad.\ Nauk.\ SSSR, 243, 1306

\reference Learned, J., \& Mannheim, K. 2000, Ann.\ Rev.\ Nucl.\ Part.\ Sci., 50, 679

\reference Lemoine, M., \& Pelletier, G. 2003, ApJ, 589, 73

\reference Linsley, J. 1975, Phys.~Rev.~Lett., 34, 1530

\reference Malkan M.A., \& Stecker F.W, 1998, ApJ 496, 13

\reference Mannheim K. 1993, A\&A, 269, 67

\reference Mannheim K. 1995,  Astropart.\ Phys., 3, 295

\reference Mannheim, K., Protheroe, R.J., \& Rachen, J.P. 2001, Phys.\ Rev.\ D, 63, id. 023003 

\reference Mastichiadis, A., Protheroe, R.J. \& Szabo, A.P. 1994, MNRAS, 266, 910

\reference Medina Tanco, G.A. 1998, ApJ, 505, L79

\reference Meli, A., \& Quenby, J., 2003, Astropart.\ Phys.,  19, 637

\reference Meli, A., \& Quenby, J., 2003, Astropart.\ Phys., 19,  649

\reference M\"{u}cke, A., Engel, R., Rachen, J.P., Protheroe, R.J., \& Stanev, T., 2000, Comp.\ Phys.\ Com., 124, 290

\reference M\"{u}cke, A. \& Protheroe, R.J. 2001, Astropart.\ Phys., 15, 121
M\"{u}cke, A., Protheroe, R.J., Engel, R., Rachen, J.P., \& Stanev, T. 2003, Astropart.\ Phys., 18, 593 

\reference M\"uller, D., Swordy, S.P., Meyer, P., L'Heureux, J., \& Grunsfeld, J.M., 1991, ApJ, 374, 356 

\reference Nagano, M., \& Watson, A.A. 2000, Rev.\ Mod.\ Phys., 72, 3, 689

\reference Norman, C.A., Melrose, D.B., \& Achterberg, A. 1995, ApJ, 454, 60

\reference Ostrowski, M. 1999, in Proc.\ Vulcano Workshop 1998: Frontier Objects in Astrophysics and Particle Physics, 1998 in Vulcano. Edited by F. Giovannelli \&   G. Mannocchi. Ital.\ Phys.\ Soc.\ Conf.\ Proc., Vol. 65. (Bologna, Italy: Italian Physical Society, 1999), p.\ 313

\reference Ostrowski, M., \& Bednarz, J. 2002, A\&A, 394, 1141

\reference P\"as, H., \& Weiler, T.J. 2001, Phys.\ Rev.\ D, 63, id. 113015

\reference Peacock, J.A. 1981, MNRAS, 196, 135

\reference Protheroe, R.J. 1986, MNRAS, 221, 769

\reference Protheroe, R.J. 1990, MNRAS, 246, 628

\reference Protheroe, R.J., Mastichiadis A. \& Dermer C.D. 1992, Astropart.\ Phys., 1, 113 

\reference Protheroe, R.J., \& Stanev, T.S. 1993, MNRAS,  264, 191

\reference Protheroe, R.J., \& Johnson, P.A. 1995, Astropart.\ Phys., 4, 253; erratum 1996,Astropart.\ Phys., 5, 215

\reference Protheroe, R.J., Stanev, T., \& Berezinsky, V.S. 1995, Phys.\ Rev.\ D, 51, 4134

\reference Protheroe R.J., \& Johnson P. 1996, Astropart.\ Phys., 4, 253

\reference Protheroe, R.J., \& Biermann, P.L. 1996, Astropart.\ Phys.,  6, 45

\reference Protheroe R.J., \& Stanev T. 1996, Phys. Rev. Lett., 77, 3708; erratum  1997, Phys.\ Rev.\ Lett., 78, 2420

\reference Protheroe R.J., 1997, in ``Accretion Phenomena and Related Outflows'', IAU Colloquium 163, ed. D.T.\ Wickramasinghe, G.V. Bicknell, \& L. Ferrario 1997, ASP Conf.\ Ser., Vol. 121, p. 585

\reference Protheroe, R.J. 1998, in 18th Int. Conf. on Neutrino Physics and Astrophysics (Neutrino 98), Takayama, Nucl.\ Phys.\ B Proc.\ Suppl.,  77,  465

\reference Protheroe, R.J., \& Stanev, T. 1999, Astropart.\ Phys., 10, 185 

\reference Protheroe, R.J., in ``Topics in cosmic-ray astrophysics'' ed. M.A.  DuVernois, Nova Science Publishing: New York, 2000, pp 258--298

\reference Protheroe, R.J. 2001, in Proceedings of XXVII Cosmic Ray Conference, Hamburg, Edited by M. Simon et al., Copernicus Gesellschaft 2001, Vol. 6, p. 2006, and ibid, Vol. 6, p. 2014

\reference Protheroe, R.J., Donea, A.-C., \& Reimer, A. 2003, Astropart.\ Physics, 19, 559

\reference Puget, J.L., Stecker, F.W., \& Bredekamp, J.H. 1976, ApJ, 205, 638

\reference Rachen, J.P., \& Biermann, P.L., 1993, A\&A, 272, 161

\reference Ressell, M.T., \& Turner, M.S. 1990, Comm.\ Astrophys., 14, 323

\reference Rubin, N. 1999, M. Phil. Thesis, University of Cambridge

\reference Sarkar, S. 2000, in COSMO-99, in Third Intern.\ Workshop on Particle Physics and the Early Universe, p.\ 77, hep-ph/0005256

\reference Sarkar, S., \& Toldra, R. 2002, Nucl.\ Phys.\ B, 621, 495

\reference Sigl G. et al. 1997, Phys.\ Lett.\ B, 392, 129

\reference Sigl, G., Lee, S., Bhattacharjee, P., \& Yoshida, S. 1999, Phys.\ Rev.\ D, 59,  id. 043504

\reference Sigl, G. 2000, Lect.~Notes~Phys., 556, 259

\reference Sigl, G., Torres, D.F., Anchordoqui, L.A., \& Romero, G.E 2001,
Phys.\ Rev.\  D, 63,  id. 081302

\reference Singh, S., \& Ma, C-P. 2003, Phys.\ Rev., D 67, 023506

\reference Smith, A.G.K., \& Clay, R.W. 1997, Aust.~J.~Phys., 50, 827

\reference Sorrell, W.H. 1987, ApJ., 323, 647

\reference Stanev, T., Biermann, P.L., Lloyd-Evans, J., Rachen, J.P., \& Watson, A.A. 1995, Phys.\ Rev.\ Lett., 75, 3056

\reference Stanev, T., Engel, R., M\"{u}cke, A., Protheroe, R.J., \& Rachen, J.P. 2000, Phys.\ Rev.\ D, 62, 093005

\reference Stecker, F.W., \& Salamon, M. 1999, ApJ, 512, 521

\reference Szabo, A.P., \& Protheroe, R.J. 1994, Astropart.\ Phys., 2, 375

\reference Takeda, M. et al.\ 2003, Astropart.\ Phys., 19, 447

\reference Tinyakov, P.G.,  \& Tkachev, I.I., 2001a,  JETP Lett., 74, 1

\reference Tinyakov, P.G.,  \& Tkachev, I.I., 2001b,  JETP Lett., 74, 445

\reference Tinyakov, P.G.,  \& Tkachev, I.I., 2003, astro-ph/0301336

\reference Tkaczyk, W., Wdowczyk, J., \& Wolfendale, A.W. 1975, J.\ Phys.\ A, 8, 1518

\reference Toshito, T.\ et al. 2001, hep-ex/0105023

\reference Totani, T. 1999, Astropart.\ Phys., 11, 451

\reference Vietri, M. 1995, ApJ, 453, 883

\reference Waxman, E., \& Bahcall, J. 1999, Phys.\ Rev.\ D, 59, id. 023002

\reference Weiler, T. 1999, Astropart.\ Phys., 11 303

\reference Zatsepin, G.T., \& Kuzmin, V.A. 1966, Sov.\ Phys.\ JETP Lett., 4, 78

\end{document}